%% file: 00_main.tex
\documentclass[conference,compsoc]{IEEEtran}
\ifCLASSOPTIONcompsoc
  \usepackage[nocompress]{cite}
\else
  \usepackage{cite}
\fi

\newcommand{\keyword}[1]{\textsf{\small #1}}

\renewcommand{\_}{\discretionary{\textunderscore}{}{\textunderscore}}

\makeatletter
\newcommand\myparagraph{%
  \@startsection{paragraph}{4}%
    {\z@}
    {1.5ex \@plus 1ex \@minus .2ex}
    {-0.5em}
    {\normalfont\normalsize\bfseries}}
\makeatother

\newcommand{\shortsection}[1]{\noindent\myparagraph{\bf #1}} 

\makeatletter
\newcommand\shortersection{%
  \@startsection{paragraph}{4}%
    {\z@}
    {1.5ex \@plus 1ex \@minus .2ex}
    {-0.5em}
    {\normalfont\normalsize\it}}
\makeatother

\usepackage[T1]{fontenc}
\newcommand{\pgir}{\textit{PGIR}\xspace}
\newcommand{\op}[1]{\textsc{#1}}

\usepackage{listings}
\usepackage{multirow}
\usepackage{makecell}
\usepackage[ruled,vlined]{algorithm2e}
\usepackage{float}
\usepackage{url}
\usepackage[hidelinks]{hyperref}
\usepackage{booktabs}
\usepackage{graphicx}
\usepackage{amsmath}
\usepackage{tabularx}
\usepackage{xcolor}
\usepackage{threeparttable}

\usepackage{enumitem}
\setlist{nosep,topsep=2pt,parsep=0pt,itemsep=2pt}

\newif\ifARXIVVERSION
\newif\ifCAMERAREADY

\CAMERAREADYfalse
\ARXIVVERSIONtrue

\begin{document}

\urlstyle{sf}

\title{Evolution of Log-Based Detection Rules in Public Repositories} 

\ifARXIVVERSION
\author{
  Minjun Long \\
  University of Virginia \\
  \and
  David Evans \\
  University of Virginia
}
\fi

\maketitle

\begin{abstract}
While prior work has examined the evolution of network intrusion detection signatures, the longitudinal behavior of log-based detection rules has received little empirical study even though
log-based detection rules remain central to modern security operations. These rules encode domain expertise that analysts iteratively refine to balance detection coverage against alert volume. We present the first longitudinal analysis of detection rule evolution across two widely used repositories: the community-driven Sigma project and the curated Splunk Security Content (SSC). To compare rule versions based on detection logic rather than surface syntax, we introduce a predicate graph intermediate representation that canonicalizes the logical structure of a rule, together with a tree alignment procedure for analyzing changes across revisions. We apply this method to 6,859 rule histories from Sigma and SSC and find that roughly 56\% of rules undergo at least one revision on detection logic.  Across rule lifetimes, evolution is predominantly non-monotonic, with over half of rules both adding and removing clauses over time. 
Combining structural analysis with LLM-based inference and human validation of operational intent shows that roughly a quarter to a third of rules alternate between expanding coverage and reducing false positives, rather than converging toward a stable form. Our analysis reveals that detection rule evolution in public repositories reflects ongoing operational trade-offs rather than steady convergence. Our study raises questions about why rules change the way they do and argues for research towards better processes for devising and deploying security rules.
\end{abstract}

\input{01_introduction}

\input{02_related_work}
\input{03_dataset}
\input{04_methodology}
\input{05_temporal_analysis}
\input{06_structural_changes}

\input{07_intent_analysis}
\input{08_discussion}

\ifARXIVVERSION 
\subsection*{Acknowledgments}
\noindent
This work is supported in part by funds provided by the National Science Foundation, Department of Homeland Security, and IBM through the ACTION AI Institute (Award \#2229876).
\fi

\bibliographystyle{IEEEtran}
\bibliography{reference}

\appendices

\input{appendix}
\ifCAMERAREADY
\clearpage 
\input{meta_review}
\fi

\end{document}

%% file: 01_introduction.tex
\section{Introduction}
\label{sec:introduction}

Host-based intrusion detection systems (HIDS) remain critical to enterprise security operations and log-based detection rules play a central role in modern Security Information and Event Management (SIEM)-driven security operations centers (SOCs). 
Despite widespread interest in anomaly detection and machine learning, rules continued to be widely used because they encode interpretable domain expertise and can be incrementally adapted to balance coverage and false positives under changing environments.

Although longitudinal studies have been conducted for signatures in network intrusion detection systems (NIDS) \cite{rulingtherules}, no similar analysis has previously been done for log-based detection rules despite their critical role in enterprise security.
%
\hyphenation{pro-ject}
Access to enterprise rules is typically restricted by organizational policies, but public rule repositories provide an accessible entry point for building understanding of how host-based rules evolve. Splunk's open-source Security Content (SSC) and the community-maintained Sigma project encode rich evolution traces—commits, diffs, metadata, and release histories—that capture multi-year histories of rule development.
While these repositories cannot directly measure  enterprise-specific tuning burdens, they expose adjustments such as added conditions, refined field usage, and structural refinements that mirror the same pressures SOC analysts describe when triaging false positives and maintaining coverage~\cite{alahmadi2022falsepositives}.
By studying these host-based detection rule repositories, we seek to understand when rules change in ways that alter detection behavior, how those changes manifest structurally, and what operational pressures they reflect such as false-positive reduction or coverage expansion. 


\shortsection{Contributions}
%
Our work provides a longitudinal, structure-aware analysis of how host-based, log-driven detection rules evolve over time. We introduce a method for systematically analyzing evolution of such rules and report our findings applying this method to the Sigma and  Splunk Security Content rule repositories. 

\shortersection{Methodology {\rm (\autoref{sec:methodology})}}
We convert Sigma rules into Splunk-style searches to match the format in the Splunk content, and process them into a unified representation that captures the logical structure of detection conditions. 
%
Our predicate graph intermediate representation isolates a rule's detection logic, allowing us to compare rules across renames and reorganizations, distinguish behavior-altering changes from maintenance edits, and analyze how detection logic evolves longitudinally. 
We develop a tree alignment algorithm that enables semantic comparison of rule versions and pair this structural analysis with an LLM-based classification step that annotates each adjacent-version pair with a semantic direction and operational rationale. 
We release the end-to-end pipeline as open source code to support follow-on empirical studies.

\shortersection{Key Findings}
We apply our method to nine years (2017--2026) of commit history across the Sigma 
and SSC repositories.  
We find that roughly 56\% of rules undergo at least one revision on detection logic, and the two repositories follow different maintenance regimes (\autoref{sec:temporal}). Structural edits are coordinated rather than isolated: 42\% of predicate-changing steps carry multiple co-occurring operations, and activity concentrates on conjunctive scopes (\autoref{sec:structural}). About a third of rules swing between expanding coverage and reducing false positives across their history, and a substantial share carry unresolved precision--coverage tensions rather than converging to a stable form (\autoref{sec:intent_analysis}).

\ifARXIVVERSION
\shortersection{Availability} The complete end-to-end analysis pipeline used to produce every quantitative result, table, and figure in this paper is released in a public github repository under an open source license, available at:
\begin{center}\small
\url{https://github.com/Elena6918/Evolution-of-Log-Based-Detection-Rules}
\end{center}
\fi




%% file: 02_related_work.tex
\section{Related Work}
\label{sec:relatedwork}
We survey prior work studying host-based intrusion detection and rule management, and review longitudinal studies of network intrusion detection rulesets. 

\shortsection{Host-based Intrusion Detection}
Research on host-based intrusion detection  has largely emphasized anomaly detection and machine learning over host telemetry, including system-call modeling, sequence- and graph-based methods, and deep learning on Windows Event Logs and Linux audit logs \cite{warrender1999detecting, du2017deeplog, kim2016lstm}. While these approaches aim to automate detection, rule-based systems remain dominant in practice. 
Apruzzese et al.\ show that evaluations of learning-based intrusion detection systems often fail to reflect enterprise deployment constraints, limiting practical adoption~\cite{apruzzese2023sok}. 
Other work examines robustness, evasion, and optimization of SIEM and Sigma rules~\cite{siem-evasion,rulegenie}. These efforts are complementary to our goal of understanding how rule logic evolves over time. In practice, SOCs remain heavily dependent on rule-based detection, where interpretability and contextual tuning are essential. This motivates our work to study how detection logic is iteratively refined over time. 

\shortsection{Operational rule management and alert fatigue}
A parallel line of work addresses the operational burden of rule-based systems, particularly alert fatigue. Several works focus on correlating and contextualizing alerts to reduce analyst workload \cite{julisch2002mining, roy2026toward, sheeraz2024revolutionizing}. For example, DeepCASE~\cite{van2022deepcase} learns relationships among alerts to group related events into higher-level narratives while preserving interpretability. This line of work treats alerts as the primary object of analysis, improving how rule outputs are consumed. In contrast, we focus on the rules themselves, characterizing how detection logic evolves over time, revealing the structural changes that encode false-positive mitigation and coverage refinement.

\shortsection{Longitudinal analyses of NIDS rulesets}
Rule evolution has been studied extensively for network intrusion detection systems (NIDS). Prior work quantifies ruleset growth, alert concentration, and incident linkage in systems such as Snort and Suricata~\cite{rulingtherules}. Other studies examine rule quality and redundancy, analyzing how modifications affect detection performance and false positives~\cite{guide2023characterizing}, and proposing principles for constructing high-value rules~\cite{teuwen2025ruling}. A complementary line of work explores automatic rule generation, including semantic-aware signatures from honeynet traffic~\cite{yegneswaran2005nemean}, anomaly-driven HTTP signature generation~\cite{garcia2015automatic}, and hybrid approaches for deriving Snort rules~\cite{jaw2022hybridsnort}. In contrast, comparable longitudinal analyses for host-based detection rules are largely absent. Ours is the first study to longitudinally characterize the evolution of host-based, log-driven rules, examining how their detection logic changes over time beyond surface-level repository edits.

%% file: 03_dataset.tex
\section{Datasets}
\label{sec:dataset}

We study complete version histories of rules from two public repositories: the Sigma project and Splunk’s Security Content (SSC). For reproducibility, we fix snapshots as of 10 April~2026 and examine the full rule histories available in the repositories. 
Sigma reflects rule evolution from initial introduction starting with the first commit in December 2016, whereas SSC was open-sourced in December 2018 after a period of internal development, so early commits may not correspond to original creation. To enable consistent analysis, we convert Sigma rules to Splunk Processing Language (SPL) using Sigma's conversion tool \cite{sigma_cli}. 

\begin{figure*}[tb]
\centering
\includegraphics[width=1.0\textwidth]{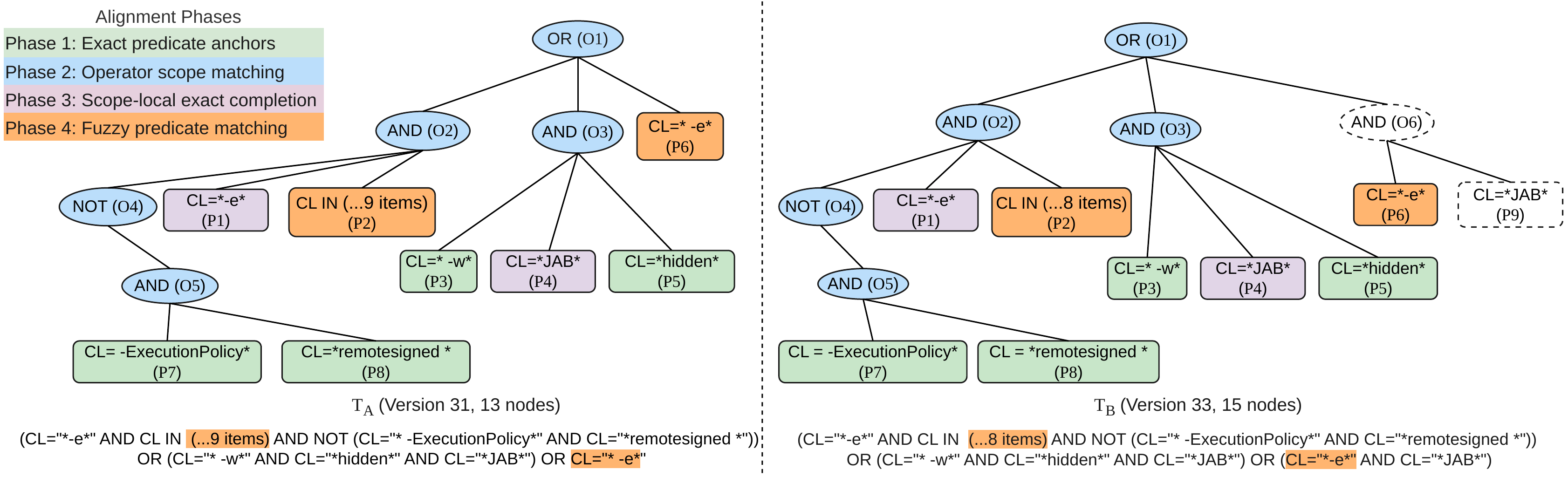}
\caption{Alignment of two versions of a PowerShell encoded-command detection rule. \rm Matched nodes share labels across versions (e.g., \keyword{\scriptsize O1}$\leftrightarrow$\keyword{\scriptsize O1},
\keyword{\scriptsize  P3}$\leftrightarrow$\keyword{\scriptsize  P3}); node color encodes which alignment phase produced each match. Dashed edges indicate unmatched insertions in $\mathrm{nodes}(T_B)$: operator \keyword{\scriptsize O6} and predicate \keyword{\scriptsize  P9} form a new \keyword{\scriptsize  AND(*-e*, *JAB*)} branch that tightens a previously unconstrained \keyword{\scriptsize  OR} arm.}
\label{fig:alignment-example}
\end{figure*}

\subsection{Rule Lineages}

To support longitudinal analysis, we organize rules into \emph{lineages}, each representing a sequence of versions intended to express the same underlying detection purpose over time.
This abstraction accommodates repository-level inconsistencies such as renaming, relocation, and identifier changes, and serves as the unit of analysis throughout the paper.
Rule evolution is not strictly one-to-one.
When a rule is split, the most semantically similar descendant continues the lineage, while others initiate new lineages.
When multiple rules are merged, the closest predecessor is treated as the continuation, and the others terminate.
This preserves continuity for measurement without requiring exact reconstruction of development history.
Applying this abstraction yields 4{,}204 lineages in Sigma and 2{,}655 in SSC, summarized in \autoref{tab:dataset_overview}. 
Sigma contains more lineages overall but undergoes fewer revisions per rule (median 9 vs.\ 39 commits in SSC). The maximum observed lifetimes correspond to the oldest active rules in each repository, with initial commits of 24~December~2016 (Sigma) and 18~December~2018 (SSC).


\begin{table}[tb]
\centering
\small
\setlength{\tabcolsep}{4.5pt}
\caption{Rule lineages studied. {\rm For active lineages, lifetime is computed up to the repository snapshot date.}}
\label{tab:dataset_overview}
\begin{tabular}{lrrrrr}
\toprule
\textbf{Dataset} 
& \textbf{Total} 
& \textbf{Active} 
& \textbf{Del.} 
& \textbf{Commits} 
& \textbf{Max life} \\
\midrule
Sigma  
& 4,204 & 3,921 & 283 & 44,565 & 3394 days \\
SSC
& 2,655 & 2,310 & 345 & 110,763 & 2669 days\\
\bottomrule
\end{tabular}
\end{table}

\subsection{Detection Rules}
\label{subsec:rules-bg}

A SIEM detection rule specifies a pattern of interest over event logs. In SPL, a rule is a pipeline of stages separated by \keyword{|}, where each stage consumes the events produced by the previous one. \emph{Filtering stages} restrict the event set by evaluating Boolean predicates, while \emph{non-filtering stages} transform, aggregate, or format results without changing which events match.
The example below (from SSC \keyword{linux\_auditd\_sudo\_or\_su\_execution}) illustrates both roles: 

\begin{lstlisting}[basicstyle=\sffamily\footnotesize,
  breaklines=true, breakatwhitespace=true, columns=flexible,
  aboveskip=4pt, belowskip=4pt]
sourcetype="auditd" proctitle IN ("*sudo *", "*su *")
| rename host as dest
| stats count min(_time) as firstTime max(_time) as lastTime BY proctitle dest
| convert timeformat="%Y-%m-%dT%H:%M:%S" ctime(firstTime)
| convert timeformat="%Y-%m-%dT%H:%M:%S" ctime(lastTime)
\end{lstlisting}

\noindent
The leading search is a filtering stage: it selects auditd events whose \keyword{proctitle} matches either \keyword{*sudo *} or \keyword{*su *}.  We model detection logic as a Boolean combination of atomic predicates of the form (\textit{field}, \textit{operator}, \textit{value}). In this example, \keyword{sourcetype="auditd"} corresponds to (\textit{sourcetype}, \textit{EQ}, \textit{auditd}), and \keyword{proctitle IN ("*sudo *", "*su *")} to a membership predicate over wildcard patterns. These predicates are combined with an implicit \keyword{AND} to define the match condition.

Our analysis captures only these filtering predicates---the components that determine which events match a rule.
The subsequent stages (\keyword{rename}, \keyword{stats}, and \keyword{convert}) aggregate and format the matched events.
These do not affect the matched event set and are not modeled in our analysis. The corresponding predicate graph representation for this example rule is shown in \autoref{appendix:pgir-example}.

%% file: 04_methodology.tex
\section{Method}
\label{sec:methodology}

We aim to characterize how detection rules evolve at the level of \emph{predicate logic}.
Rule revisions combine heterogeneous edits: value tuning, predicate addition and removal, and Boolean restructuring. This makes direct textual comparison unreliable since superficial differences (operand reordering, nested expressions) can obscure logical changes, while small edits may reflect substantial restructuring.
Exact semantic equivalence is infeasible due to heterogeneous predicate operators (e.g., wildcards, regex, membership) over unknown event domains.

We therefore design a comparison pipeline around a stable structural representation we call \emph{predicate graph intermediate representation} (\pgir) (\autoref{subsec:pgir}), which captures the essential semantics of the filtering stage of a rule to allow for structural comparisons.
\autoref{subsec:alignment} explains how we align two rule variants in their canonicalized PGIR representations, and 
\autoref{subsec:cost-model} describes the cost model we use to quantify their distance.

\subsection{Predicate Representation (\pgir)}
\label{subsec:pgir}

The goal of \pgir\ is to provide a representation of rules that preserves the detection logic while remaining invariant to syntactic variations that do not reflect meaningful semantic change. We ignore transformation, aggregation, and presentation stages to focus our comparison on detection logic rather than output shaping. Our normalization preserves semantics: any two rule versions that collapse to the same \pgir must match the same events. The converse does not hold: semantically equivalent versions may still differ structurally after canonicalization. Thus, this representation enables us to use structural differences as a conservative proxy for semantic change. 

We encode the filtering stage as an AST-like directed acyclic graph.
Internal nodes represent Boolean operators (\keyword{AND}, \keyword{OR}, \keyword{NOT}); leaf nodes represent atomic predicates of the form (\textit{field}, \textit{operator}, \textit{value}), annotated with polarity induced by negation context. Construction applies two structural normalizations that do not change meaning: associative flattening of nested \keyword{AND}/\keyword{OR} scopes, and canonical ordering of children under commutative operators. These eliminate residual regrouping and operand-permutation artifacts so that the downstream alignment algorithm can match on semantic content rather than surface form. 

\subsection{Aligning Canonical Predicate Trees}
\label{subsec:alignment}
After canonicalization, we align predicate trees to identify structurally plausible correspondences between rule versions.
Instead of a minimal edit script, we construct a partial injective mapping that aligns preserved structure while leaving true edits unmatched.
We illustrate our alignment algorithm on a concrete example next; the full pseudocode is given in Appendix~\ref{appendix:align-algo}.

\autoref{fig:alignment-example} shows two consecutive versions of a Sigma rule detecting PowerShell encoded-command abuse: $\mathrm{T}_\mathrm{A}$ (version~31, 13~nodes) and $\mathrm{T}_\mathrm{B}$ (version~33, 15~nodes).
The two rule versions share most of their logical structure, but $\mathrm{T}_\mathrm{B}$ introduces a new \keyword{AND(*-e*, *JAB*)} branch, removes one entry from an \keyword{IN}-list, and adjusts a minor whitespace variant in one value.
Alignment should match the preserved structure so that these three changes are identified. 

The alignment algorithm proceeds in four phases, each operating on the partial mapping $\Phi$ left by the previous phase.
Phases~1 and~3 extend $\Phi$ by matching predicate leaves under increasingly relaxed uniqueness conditions; Phase~2 uses the resulting predicate evidence to infer operator-node correspondences bottom-up; Phase~4 adds conservative fuzzy near-matches for structurally compatible but value-variant predicates, after which Phase~2 is re-executed to recover any operator matches newly supported by fuzzy evidence.
A match is added only when it is consistent with all prior matches---a node $i \in \mathrm{T}_\mathrm{A}$ may be mapped to $j \in \mathrm{T}_\mathrm{B}$ only if the nearest already-matched ancestor of $i$ is mapped to an ancestor of $j$. 

\shortersection{Phase~1: Global exact predicate anchors.}
Each predicate leaf is characterized by its field, operator, normalized value, and polarity.
Phase~1 matches leaves whose combined key is unique in each tree, producing high-confidence \emph{anchors} (the green nodes in \autoref{fig:alignment-example}).
Restricting to globally unique keys is critical: predicates such as \keyword{*JAB*} and \keyword{*-e*} appear in multiple branches and must not be matched arbitrarily, as doing so could misidentify which copy was inserted and which was preserved.

\shortersection{Phase~2: Bottom-up operator scope matching.}
With predicate anchors in place, operator nodes are matched by measuring how many of their anchored descendants already correspond under $\Phi$.
Candidate pairs with the same operator label are scored by the fraction of shared anchor evidence relative to both subtrees, and accepted when this agreement is sufficiently high (threshold-based) on both sides. Operators are processed bottom-up so that inner scopes are matched before the scopes that contain them. The matching operator nodes are shown in blue circles in \autoref{fig:alignment-example}.

\shortersection{Phase~3: Scope-local exact completion.}
Some predicates are globally ambiguous (appearing more than once) yet locally unambiguous once their enclosing operator scope is known.
For each already-matched operator pair, unmatched leaves are collected and any key appearing the same number of times in both subtrees is resolved by sorted order within the scope (the purple nodes in \autoref{fig:alignment-example}).

\shortersection{Phase~4: Conservative fuzzy predicate matching.}
The final pass handles near-identical leaves that differ only in minor value edits. Still-unmatched leaves in $\mathrm{nodes}(\mathrm{T}_\mathrm{A})$ are paired with candidates in $\mathrm{nodes}(\mathrm{T}_\mathrm{B})$ sharing the same operator class and value type, filtered by a similarity threshold, and selected by a score that jointly considers value similarity, scope compatibility, and operator agreement (the two orange nodes for each tree in \autoref{fig:alignment-example}).
In the figure, this recovers the whitespace-variant \keyword{*~-e*} vs.\ \keyword{*-e*} pair and the \keyword{IN}-list that lost one entry.
Phase~2 is then re-run with all matched predicates as evidence to pick up any operator matches newly supported by fuzzy pairs (not applicable in this example).

\shortsection{Result}
The mapping $\Phi$ is intentionally conservative: stable structure is aligned and genuine edits are left unmatched.
In the figure example, all 13 nodes of $\mathrm{nodes}(\mathrm{T}_\mathrm{A})$ are matched, while two nodes in $\mathrm{nodes}(\mathrm{T}_\mathrm{B})$ remain unmatched: the inserted \keyword{AND} operator and its \keyword{*JAB*} child. This precisely identifies the new tightened branch and predicate refinement.

\subsection{Predicate-Logic Change Cost Model}
\label{subsec:cost-model}

Given the alignment $\Phi$, we compute a weighted pred\-i\-cate-logic distance score between rule versions. Our formulation uses a tree edit distance framework~\cite{zhang1989}, in which dissimilarity between two trees is the minimum-cost sequence of node insertions, deletions, and relabelings that transforms one into the other. Rather than uniform
unit costs, we assign operation-specific weights that reflect the semantic weight of each edit in the detection-rule setting—an approach that parallels weighted AST differencing for source-code evolution \cite{falleri2014}, where per-operation costs are tuned to distinguish superficial refactorings from substantive behavioral changes.

\shortsection{Edit costs}
Aligned predicate leaves contribute update costs; unmatched leaves contribute insertion or deletion costs. Differences between aligned predicates are decomposed into field, operator, and value changes, with update cost capped by deletion plus insertion to prevent over-penalization. Predicate insertions and deletions reflect atomic condition changes (cost $1.0$). Within an aligned predicate, field, operator, and value-payload updates cost $0.2$, $0.5$, and $0.8$ respectively, with value updates slightly cheaper than insertions and deletions to reflect that they often represent tuning. Boolean operator edits carry substantially higher costs ($3.0$ for insert/delete, $4.5$ for label updates) to reflect structural rewrites of Boolean logic. 
Although the exact costs assigned to each type of edit are somewhat arbitrary, their relative values are sufficient to distinguish structural changes from local predicate tuning. Statistics are robust to small variations in these weights. 

%% file: 05_temporal_analysis.tex
\section{Temporal Evolution of Detection Rules}
\label{sec:temporal}

Using the method from \autoref{sec:methodology}, we study how predicate-level detection logic evolves over time in the Sigma and SSC rulesets. Our analysis operates at two levels. We use the full set of rule lineages to characterize rule creation and the prevalence of predicate changes, and we analyze changes between consecutive commits to understand how detection logic is modified over time.
Of the 146{,}566 revision steps across both corpora (\autoref{sec:dataset}), most are metadata, formatting, or pipeline changes. We analyze only predicate-changing revisions (8{,}234 in Sigma and 4{,}668 in SSC). Step-based analyses use the 3{,}942 (Sigma) and 2{,}577 (SSC) \emph{step-eligible lineages} with at least two versions.




\subsection{Rule Creation and Maintenance}
\label{subsec:temporal-dynamics}

\begin{figure}[tb]
\centering
\includegraphics[width=\columnwidth]{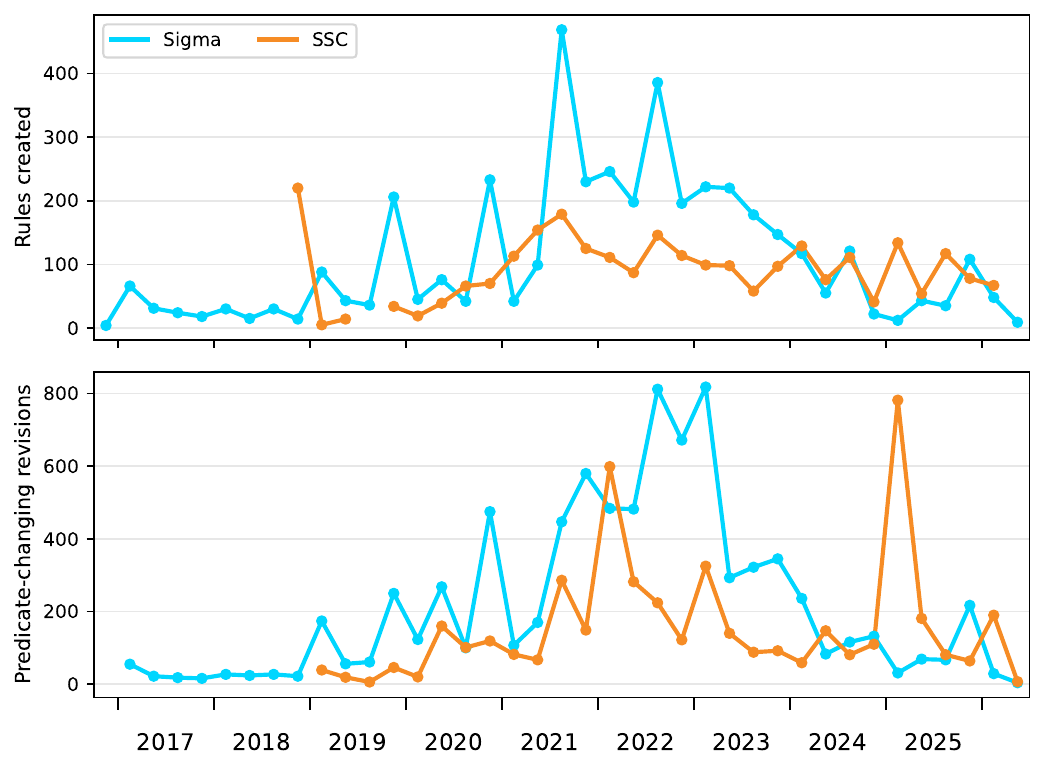}
\caption{Quarterly rule creation count (above) and revision volume (below) for Sigma (blue) and SSC (orange).}
\label{fig:temporal-revision}
\end{figure}
\begin{table}[tb]
\centering
\small
\caption{Summary of rule revisions. \rm
Prevalence is computed over all rules; other rule-level statistics over edited rules; and magnitude statistics over predicate-changing revisions.}
\label{tab:temporal-summary}
\begin{tabular}{p{5cm}rr} 
\toprule
\textbf{Metric} & \textbf{Sigma} & \textbf{SSC} \\
\midrule
Total rules                    & 4,204  & 2,655 \\
Revision steps                 & 39,480 & 107,086 \\
Predicate-changing revisions  & 8,234  & 4,668 \\[1ex]

\multicolumn{3}{c}{\emph{Prevalence (Rule-Level)}} \\[1ex]
Step-eligible rules ($\geq 2$ versions) & 3,942  & 2,577 \\
Edited ($\geq 1$ predicate change)      & 2,355  & 1,493 \\
Proportion edited                       & 56.0\% & 56.2\% \\[1.5ex]

\multicolumn{3}{c}{\emph{Predicate-Changing Revisions per Edited Rule}} \\[1ex]
Mean   & 3.5 & 3.1 \\
Median & 2   & 2 \\
90\textsuperscript{th} percentile & 7 & 7 \\[1.5ex]

\multicolumn{3}{c}{\emph{Timing of Revisions}} \\[1.5ex]
Days to first revision (median) & 60 & 148 \\
At least one revision in first 90 days & 54.4\% & 42.7\%  \\
First revision after 2 years & 8.0\% & 14.4\% \\[1.5ex]

\multicolumn{3}{c}{\emph{Revision Magnitude ($d_{\mathit{pred}}$)}} \\[1.5ex]
Mean & 5.3 & 5.5 \\
25\textsuperscript{th} percentile & 0.8 & 1.0 \\
Median & 2.0 & 2.0 \\
90\textsuperscript{th} percentile & 13.0 & 13.0 \\
Maximum & 359.0 & 564.0 \\
\bottomrule
\end{tabular}
\end{table}

\begin{figure}[tb]
  \centering
  \includegraphics[width=\columnwidth]{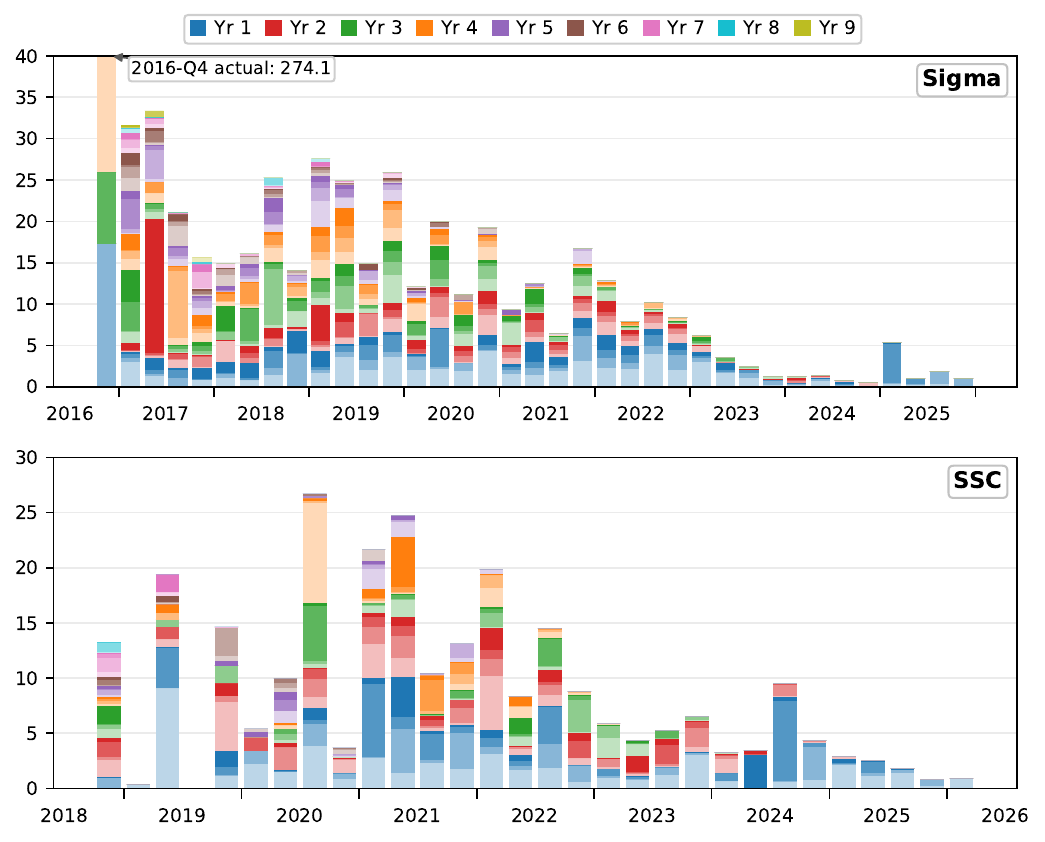}
\caption{
    Cohort-wise revisions. \rm 
    Each bar shows the average accumulated edit magnitude per rule for a creation cohort, with stacked segments indicating contributions from different lags (shaded by quarter within each year as colored). Across both repositories, accumulation is front-loaded but persists over long horizons.
}
  \label{fig:cohort-magnitude}
\end{figure}

\autoref{fig:temporal-revision} shows the counts of newly created rules and the number of predicate-changing revisions for both repositories by quarter over the lifetime of each dataset.
In Sigma, rule creation and predicate-changing revision volume rise together from 2019 through 2022, with revision activity peaking near 2022 and declining sharply thereafter despite a large rule base. SSC's creation grows more gradually without a single expansion phase, and revisions remain spread across the entire timeline with intermittent spikes including a prominent surge around 2025. 

\autoref{tab:temporal-summary} summarizes the predicate-changing revision statistics. 
Revision counts here measure only predicate-changing revisions. They track maintenance activity, but isolate effort directed toward modifying detection behavior rather than metadata or other non-semantic changes.
%
About 56\% of rules undergo at least one predicate-changing revision over their lifetime in both repositories, and among these edited rules, the median number of revisions is two (mean 3.5 in Sigma, 3.1 in SSC).

Revisions persist long after creation: 8.0\% of edited rules in Sigma and 14.4\% in SSC do not see their first predicate change until more than two years post-introduction. Sigma's initial refinement also begins much sooner (median time to first revision is 60 days, 148 for SSC) quantifying the contrast visible in \autoref{fig:temporal-revision}.
%
Edit magnitudes are heavy-tailed: median $d_{pred}=2.0$ and 90\textsuperscript{th} percentile is $\approx 13.0$ in both repositories. The largest edits are typically driven by representation changes, most commonly rewriting long disjunctions into \keyword{IN}-list predicates, which result in many predicate deletions under the cost model. Overall, the distribution indicates that rule evolution is dominated by incremental adjustments with occasional substantial restructuring.

\shortsection{Rule cohorts}
\label{subsubsec:cohort}
\autoref{fig:cohort-magnitude} aggregates predicate-change magnitude per creation cohort by the lag time between creation and revision. The most immediate observation is heterogeneity: Sigma's earlier cohorts (2016--2019) accumulate substantially more lifetime edit magnitude per rule than later ones, while SSC is broadly flatter with isolated high-magnitude exceptions.

We note two cohort outliers. Sigma's 2016 Q4 cohort (3 rules; bar clipped at 40, actual value 274.1) is dominated by \keyword{win\_alert\_mimikatz\_keywords}, which detects command-line invocations of Mimikatz (a widely-used credential-dumping tool) by matching against a curated keyword list (\autoref{appendix:mimikatz-history} provides the full timeline). SSC's 2020 Q3 cohort is similarly distorted by one rule, \keyword{ssa\_\_\_system\_process\_running\_from\_unexpected\_location}, but through prolonged structural reformulation including an SPL2 conversion (v76$\rightarrow$v77, $d_{pred}=564.0$) before eventual deletion. 
The two outliers reflect distinct mechanisms: a tiny-cohort restructuring burst for the Sigma outlier versus prolonged instability in a complex lineage for the SSC outlier.

\subsection{Representative Temporal Patterns}
\label{subsec:temporal-representative}

The aggregate statistics above characterize when revisions occur across the corpus, but do not reveal how these revisions unfold within individual rules. To provide a more concrete view, \autoref{tab:maintenance-archetypes} classifies each rule by the windows in which it received at least one predicate-changing edit: the creation quarter ($\ell_0$), the remainder of the first two years ($\ell_{1\text{--}7}$), or beyond two years post-creation ($\ell_{\geq 8}$). To control for right-censoring, we restrict to rules with at least three years of observable lifetime (3{,}251 of 4{,}204 Sigma rules and 2{,}086 of 2{,}655 SSC rules).

Late maintenance is common: 30.1\% of Sigma rules and 32.1\% of SSC rules receive at least one edit more than two years after creation. Despite differences in overall revision volume (\autoref{subsec:temporal-dynamics}), the prevalence of late edits is similar across the two corpora.

Edits are often concentrated in specific windows rather than uniformly distributed over time. Single-window patterns account for a substantial fraction of rules in both datasets, particularly mid-only edits ($(0,\geq\!1,0)$: 17.6\% Sigma, 17.4\% SSC) and creation-only edits ($( \geq\!1,0,0)$: 9.0\% vs.\ 8.0\%).
Finally, multi-window maintenance remains common, with 34.3\% of Sigma rules and 25.7\% of SSC rules receiving edits across multiple stages of their lifetime. Within this group, mid+late ($(0,\geq\!1,\geq\!1)$) and all-window ($( \geq\!1,\geq\!1,\geq\!1)$) patterns indicate continued refinement beyond initial deployment.

\begin{table}[tb]
\centering
\small
\caption{Lifecycle archetypes. \rm Rules with at least three years of observed lifetime (3{,}251 Sigma, 2{,}086 SSC) 
by number of edits in the creation quarter ($\ell_0$), next two years ($\ell_{1-7}$), and thereafter ($\ell_{\ge 8}$).}
\label{tab:maintenance-archetypes}
\begin{tabular}{lrr}
\toprule
\textbf{Archetype} & \textbf{Sigma} & \textbf{SSC} \\
\midrule
Never edited $(0,0,0)$ 
  & 1{,}065 (32.8\%) & 787 (37.7\%) \\
\midrule
\multicolumn{3}{l}{\emph{Single-window}} \\
\quad Creation-only $(\geq\!1, 0, 0)$ 
  & 294 (9.0\%)  & 167 (8.0\%) \\
\quad Mid-only $(0, \geq\!1, 0)$ 
  & 572 (17.6\%) & 363 (17.4\%) \\
\quad Late-only $(0, 0, \geq\!1)$ 
  & 206 (6.3\%)  & 233 (11.2\%) \\
\midrule
\multicolumn{3}{l}{\emph{Multi-window}} \\
\quad Creation + Mid  $(\geq\!1, \geq\!1, 0)$ 
  & 341 (10.5\%) & 99 (4.7\%) \\
\quad Creation + Late $(\geq\!1, 0, \geq\!1)$ 
  & 131 (4.0\%)  & 51 (2.4\%) \\
\quad Mid + Late      $(0, \geq\!1, \geq\!1)$ 
  & 387 (11.9\%) & 278 (13.3\%) \\
\quad All three       $(\geq\!1, \geq\!1, \geq\!1)$ 
  & 255 (7.8\%)  & 108 (5.2\%) \\
\bottomrule
\end{tabular}
\end{table}

\shortsection{Case study: a late-only SSC rule}
Among the archetypes, late-only rules are the most analytically informative. They survive the typical refinement window untouched, so the eventual revision must be driven by something other than initial settling. They are also the archetype where Sigma (6.3\%) and SSC (11.2\%) diverge most sharply, making them a natural lens on the long-horizon maintenance behavior already visible at the population level (\autoref{subsec:temporal-dynamics}). 

We illustrate the archetype with an SSC rule that exhibits two distinct mechanisms in a single lineage:
\keyword{possible\_lateral\_movement\_powershell\_spawn}. This rule lay dormant for ten quarters before any revision, but was revised four times over the next seven quarters. The original form enumerates parent and child process names as flat \keyword{OR} chains with no exclusions:\footnote{In the listings, the \keyword{\scriptsize Processes} field prefix and the \keyword{\scriptsize tstats} preamble and postamble are omitted; only the \keyword{\scriptsize where} predicate is shown.}

\begin{lstlisting}[
  basicstyle=\sffamily\footnotesize,
  breaklines=true, breakatwhitespace=false, columns=flexible,
  escapeinside={(*}{*)},
  xleftmargin=0pt,
  breakindent=0pt,
  aboveskip=4pt, belowskip=4pt
]
(*\textsf{\bf v44, original (2021-Q4):}*)
(parent_process_name=wmiprvse.exe OR parent_process_name=services.exe OR parent_process_name=svchost.exe  OR parent_process_name=wsmprovhost.exe OR parent_process_name=mmc.exe) AND (process_name=powershell.exe OR (process_name=cmd.exe AND process=*powershell.exe*) OR process_name=pwsh.exe OR (process_name=cmd.exe AND process=*pwsh.exe*))
\end{lstlisting}
\noindent
The lag-10 revision (v45) leaves the detection structure intact but appends a \keyword{NOT} clause excluding processes under the SCCM client path, suppressing a recurring class of false positives from software deployment agents. Seven quarters later, the lag-17 revision restructures both predicate groups: the flat \keyword{OR} enumerations become \keyword{IN} lists, and \keyword{svchost} receives a separate sub-clause filtering out known-benign service invocation contexts (\keyword{netsvcs}, \keyword{Schedule}) rather than treating all \keyword{svchost} spawns as equal:

\begin{lstlisting}[
  basicstyle=\sffamily\footnotesize,
  breaklines=true, breakatwhitespace=false, columns=flexible,
  escapeinside={(*}{*)},
  xleftmargin=0pt,
  breakindent=0pt,
  aboveskip=4pt, belowskip=4pt
]
(*\textsf{\bf v91, lag 17 (2026-Q1):}*)
(parent_process_name IN ("mmc.exe","services.exe", "wmiprvse.exe", "wsmprovhost.exe") OR (parent_process_name="svchost.exe" NOT parent_process IN ("*-k netsvcs*","*-s Schedule*"))) AND (process_name IN ("powershell.exe","pwsh.exe") OR (process_name=cmd.exe process IN ("*powershell*","*pwsh*"))) NOT process IN ("*C:\Windows\CCM\*")
\end{lstlisting}

The trajectory illustrates two distinct drivers of late revisions: a false-positive suppression edit prompted by SCCM noise, then a structural reformulation that replaces flat \keyword{OR} enumerations with \keyword{IN} lists and adds nuanced \keyword{svchost} filtering. Other late revisions in the corpus are stylistically driven---e.g., collapsing redundant command-line templates into more compact \keyword{IN} expressions while preserving coverage. In both forms, the detection intent is preserved while the predicate representation changes substantially.

%% file: 06_structural_changes.tex
\section{Structural Evolution of Predicate Logic}
\label{sec:structural}

Section~\ref{sec:temporal} characterized \emph{when} predicate-logic changes occur and how they are distributed over time; here, we examine \emph{how} detection logic evolves structurally, focusing on the changes within predicate logic.
The weighted predicate distance of \autoref{subsec:cost-model} captures revisions as atomic insertions, deletions, and updates, but those edit primitives are too fine-grained to be directly interpretable.

\autoref{subsec:structural-operation} defines a set of \emph{structural operation labels} over canonicalized predicate trees, which aggregate edit primitives into higher-level, human-readable transformations of logical structure.
We use that taxonomy to analyze the frequency of structural operations (\autoref{subsec:structural_frequency}), finding that conjunctive additions outpace disjunctive additions by roughly 3 to 6 times. Adding an \keyword{AND} predicate accounts for 33.8\% of predicate-changing steps in SSC and 23.5\% in Sigma, compared to 6.0\% and 8.1\% for adding an \keyword{OR} predicate. 
\autoref{subsec:structural_cooccurrence} studies the co-occurrence of different structural operations, and \autoref{subsec:structural_patterns} analyzes patterns of structural evolution in rule lineages. 

At the lineage level, more than half of pred\-i\-cate-changing rules in both datasets evolve non-monotonically, mixing expansion and contraction over their lifetime. Most of this mixing is \emph{intra-step}---a single revision both adds and removes structure---but a substantial fraction unfolds across multiple revisions, including explicit structural reversions ($A \rightarrow B \rightarrow A$) that affect 25.4\% of predicate-changing lineages in SSC versus 9.2\% in Sigma.

\subsection{Structural Operation Taxonomy}
\label{subsec:structural-operation}

We define structural operation labels over predicate trees to summarize structural changes between adjacent versions.
Each revision step is represented as a \emph{set of operations}, allowing multiple co-occurring changes. Each structural operation is characterized using one of eight different labels:

\begin{description}
    \item \emph{Predicate-set modifications} (\op{and+}, \op{and-}, \op{or+}, \op{or-}):
These operations change the set of predicates within an existing Boolean scope, adding (\op{and+}, \op{or+}) or removing (\op{and-}, \op{or-}) either a predicate from a conjunctive (\op{and}) scope (including the root) or a disjunctive (\op{or}) scope.
  \item \emph{Scope introduction and removal} (\op{branch+}, \op{branch\mbox{-}}): Introduce or eliminate an entire subtree.
\item \emph{Structural reorganization} (\op{move}, \op{flip}):
A \op{move} operation relocates a matched predicate to a different enclosing Boolean scope; a \op{flip} operation changes a scope's operator type (\keyword{AND}~$\leftrightarrow$~\keyword{OR}) without relocation.
      
\end{description}
By construction, these labels cover all structural transformations between aligned predicate trees: every difference in tree shape or scope assignment surfaced by the alignment $\Phi$ corresponds to one or more of these operations. Revisions with $d_{pred} > 0$ but no structural label consist entirely of value-level updates (\op{val-update}) within predicates; we analyze these separately at the lineage level in \autoref{subsec:structural_patterns} as the \op{value-only} pattern.

\subsection{Frequency of Structural Operations}
\label{subsec:structural_frequency}

\begin{table}[tb]
\centering
\setlength{\tabcolsep}{4pt}
\small
\caption{
Distribution of structural operations. \rm
Each count cell is the proportion of predicate-changing revision steps that contain structural revisions with that number of structural operations. 
}
\label{tab:structural-coverage}
\begin{tabular}{lrccccccc}
\toprule
\textbf{Corpus} & \multicolumn{1}{c}{\textbf{Steps}} & \textbf{Structural} & \textbf{Avg} & \textbf{1} & \textbf{2} & \textbf{3} & \textbf{4} & \textbf{5+} \\
\midrule
Sigma & 8,234 & 5,241 & 1.80 & .37 & .50 & .09 & .03 & .007  \\
SSC   & 4,668 & 3,570& 1.75 & .39 & .52 & .06 & .03 & .006 \\
\bottomrule
\end{tabular}
\end{table}

Structural operation labels apply broadly across predicate-changing revisions. Overall, the majority of steps receive at least one structural label (63.7\% in Sigma; 76.5\% in SSC), while the remaining steps are value-level updates with no change to logical structure. Among steps that do carry structural edits, as shown in \autoref{tab:structural-coverage}, both average roughly 1.8 operations per step. For both rulesets, over 60\% carry two or more operations, indicating that structural changes are often compositional rather than isolated.  Having more than three operations, though, is quite rare and less than one percent of the steps involve more than four operations.  

\begin{figure}[tb]
\centering
\includegraphics[width=\columnwidth]{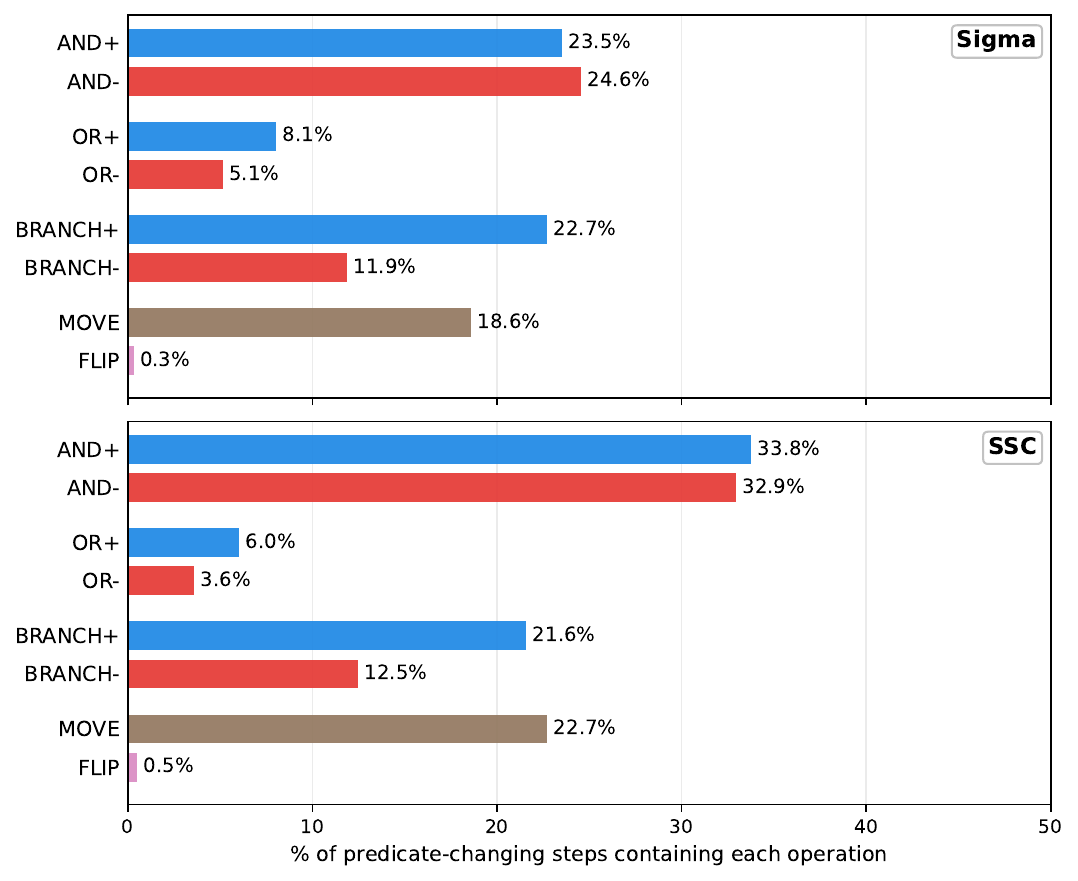}
\caption{
Prevalence of structural operations among predicate-changing revision steps. \rm Each bar shows the fraction of steps in which a given operation appears; steps with no structural operation (\op{value-only} edits) are included in the denominator but contribute to no bars.
}
\label{fig:primitive-prevalence}
\end{figure}

\autoref{fig:primitive-prevalence} shows the prevalence of each structural operation among all predicate-changing revision steps (including those with no structural operations).
Conjunctive modifications dominate in both corpora, with \op{and+} far more common than \op{or+}. This asymmetry is consistent across both datasets, but more pronounced in SSC.
We do not interpret this pattern as a direct signal of operational intent: the effect of adding an \keyword{AND} scope depends on its interaction with other edits. We treat this as a structural observation about \emph{where} edits concentrate, and defer the question of \emph{why} to \autoref{sec:intent_analysis}.

Within the conjunctive family, both corpora show near-balance between additions and removals (23.5\% \op{and+} vs.\ 24.6\% \op{and-} in Sigma; 33.8\% vs.\ 32.9\% in SSC). 
Beyond predicate-set changes, structural reorganization is also common. Branch-level edits and predicate relocation each occur in a significant fraction of steps, indicating that rule evolution frequently involves restructuring logical structure in addition to modifying predicate sets.
Finally, \op{flip} operations are rare ($<1\%$ in both corpora). When flips do occur, they are typically accompanied by other structural modifications, most commonly \op{move}, rather than appearing in isolation. Across both corpora, these events are split between \keyword{and}$\rightarrow$\keyword{or} (54.6\%) and \keyword{or}$\rightarrow$\keyword{and} (45.4\%), suggesting no strong directional bias.

\subsection{Co-occurrence of Structural Operations}
\label{subsec:structural_cooccurrence}

\begin{figure}[!t]
\centering
\includegraphics[width=1\columnwidth]{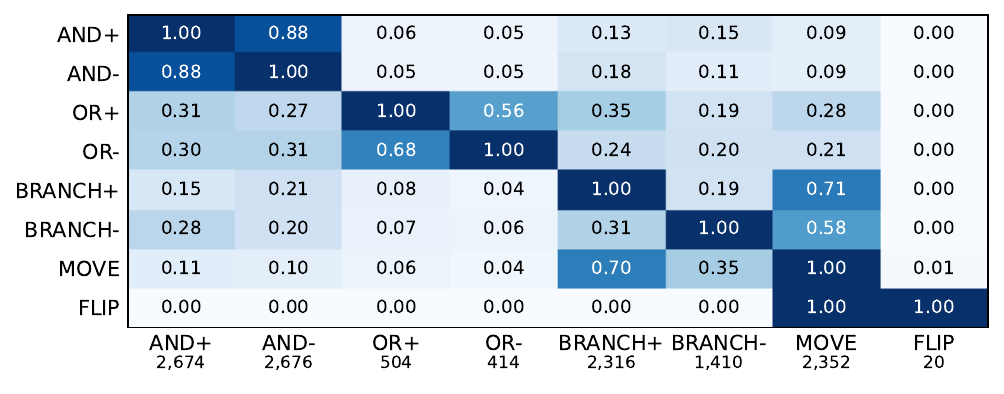}
\caption{
Co-occurrence of structural operation labels. \rm
Each cell reports $P(\text{column op} \mid \text{row op})$.
The number below each column gives the number of multi-label revision steps in which that operation appears.
}
\label{fig:structural-coocurrence-heatmap}
\end{figure}

As shown in \autoref{fig:structural-coocurrence-heatmap}, structural evolution operates through coordinated rewrites spanning predicate composition, scope assignment, and hierarchical organization, rather than by one-dimensional edits applied in isolation. Each cell in the figure reports $P(\text{column op} \mid \text{row op})$ over steps that involve multiple structural labels. 

Because multi-operation steps are common (\autoref{tab:structural-coverage}), this analysis reveals how edits combine.
The co-occurrence structure is far from random. Within existing Boolean scopes, additions and removals frequently appear together: \op{and+} and \op{and-} are strongly and almost symmetrically coupled, while \op{or+} and \op{or-} show the same pattern but less strongly. This suggests that many intra-scope edits are not simple expansions or contractions, but coordinated substitutions of one predicate set for another, consistent with shifts in detection focus.

At a higher structural level, branch introduction and removal are tightly linked to predicate relocation. \op{branch+} and \op{move} co-occur at high rates in both directions, and \op{branch-} shows the same tendency, indicating that edits to Boolean hierarchy typically involve repositioning existing pred\-i\-cates within that hierarchy rather than adding or deleting branches independently. Likewise, every \op{flip} step in this multi-label population also carries a \op{move}, suggesting that AND/OR relabeling rarely appears as a standalone connective rewrite. Instead, it tends to arise as part of broader reorganizations in which a preserved local context changes Boolean type while matched predicates are simultaneously redistributed across scopes.

\subsection{Structural Evolution Patterns}
\label{subsec:structural_patterns}

\begin{table}[tb]
\centering
\small
\caption{Structural evolution patterns. \rm Distribution across rule lineages with at least one predicate-changing revision.
}
\label{tab:structural-patterns}
 
\begin{tabular}{lr@{\hspace{0.5em}}l@{\hspace{3em}}r@{\hspace{0.5em}}l}
\toprule
\textbf{Pattern} & \multicolumn{2}{c}{\textbf{Sigma}} & \multicolumn{2}{c}{\textbf{SSC}} \\
\midrule
Number of Lineages   & \multicolumn{2}{c}{2,355} & \multicolumn{2}{c}{1,493} \\
\op{value-only}       & 451   & (19.2\%) & 118 & (7.9\%)  \\
\op{expand-only}      & 309   & (13.1\%) & 396 & (26.5\%) \\
\op{contract-only}    & 250   & (10.6\%) & 98  & (6.6\%)  \\
\op{restructure-only} & 12    & (0.5\%)  & 46  & (3.1\%)  \\
\op{mixed}            & 1,333 & (56.6\%) & 835 & (55.9\%) \\
\bottomrule
\end{tabular}
\end{table}

While structural operations characterize individual revision steps, rule evolution unfolds across sequences of revisions spanning years.
We analyze structural evolution at the level of \emph{lineages}, grouping rules based on how structural modifications accumulate over their lifetime. 

\shortsection{Structural evolution patterns}
We define mutually exclusive lineage-level patterns based on the families of struc\-tur\-al operations observed across all revisions of a rule.
These patterns are defined over syntactic transformations of predicate logic rather than their semantic effect on matched events.
Expansion operations introduce additional predicates or Boolean structure (\op{and+}, \op{or+}, \op{branch+}), while contraction operations remove them (\op{and-}, \op{or-}, \op{branch-}).
Reorganization operations (\op{move}, \op{flip}) preserve the predicate set but alter its Boolean context.
%
These categories distinguish monotonic (expansion/contraction) from non-monotonic trajectories, and separate both from value-only and reorganization-only evolution: 
\begin{description}
\item \op{value-only}: only predicate values change (no change to logical structure).
\item \op{expand-only}: only expansion operations.
\item \op{contract-only}: only contraction operations.
\item \op{restructure-only}: only reorganization operations.
\item \op{mixed}: both expansion and contraction occur.
\end{description}


As shown in \autoref{tab:structural-patterns}, \op{mixed} is the most common pattern in both corpora (56.6\% in Sigma; 55.9\% in SSC), indicating that many rules undergo both structural growth and reduction rather than evolving in a single direction. To clarify what this category captures, we further distinguish whether expansion and contraction co-occur within the same revision step (\emph{intra-step mixing}) or arise across different steps in the same lineage (\emph{inter-step alternation}). 
Most \op{mixed} lineages contain intra-step mixing, either alone (67.1\% in Sigma; 61.0\% in SSC) or together with inter-step alternation (18.2\% in Sigma; 19.6\% in SSC). Inter-step alternation without any mixed step is less common (14.8\% in Sigma; 19.4\% in SSC).
Thus, the prevalence of \op{mixed} is driven primarily by revisions that combine expansion and contraction within the same step, rather than only by back-and-forth changes across separate revisions.

Beyond this shared pattern, the two corpora differ in their distribution of monotonic trajectories.
\op{contract-only} (10.6\% in Sigma; 6.6\% in SSC) and \op{value-only} are more common in Sigma (19.2\% in Sigma; 7.9\% in SSC), while \op{expand-only} is more prevalent in SSC (26.5\% vs.\ 13.1\% in Sigma).
These differences indicate that SSC contains a larger fraction of lineages that accumulate additional structure over time, whereas Sigma includes more lineages that either simplify structure or operate primarily through value-level modifications without structural change.
The underlying causes of these differences are an open question, but these results are consistent with distinct development and maintenance practices across the two repositories observed throughout our analysis.



\begin{table}[tb]
\centering
\small
\caption{
Structural reversions (A--B--A patterns).
}
\label{tab:aba-summary}
\begin{tabular}{lcc}
\toprule
\textbf{Metric} & \textbf{Sigma} & \textbf{SSC} \\
\midrule
Number of Lineages & 2,355 & 1,493 \\
Lineages with $\geq$1 A--B--A (\%) & 216 (9.2\%) & 379 (25.4\%) \\
Total A--B--A triplets & 383 & 712 \\
Median restore time (hours) & 114.5 & 159.6 \\
Restored in $\leq$24h & 45.4\% & 22.3\% \\
Restored in $\leq$7d  & 65.5\% & 51.8\% \\
\bottomrule
\end{tabular}
\end{table}

\shortsection{Structural reversion (A--B--A patterns)}
To further characterize non-monotonic evolution, we identify exact A--B--A patterns: three consecutive versions $(v_i, v_{i+1}, v_{i+2})$ where $v_i$ and $v_{i+2}$ share the same predicate-graph structure but $v_{i+1}$ differs.
Because the triplet spans consecutive versions with no intervening revisions, the restore time ($v_{i+1} \to v_{i+2}$) measures how quickly the next commit reverses the structural change.

As shown in \autoref{tab:aba-summary}, such reversions occur in 9.2\% of all predicate-changing lineages in Sigma (216 of 2,355) and 25.4\% in SSC (379 of 1,493), totaling 383 and 712 triplets respectively.
Multiple reversions are common: 32.4\% of Sigma A--B--A lineages and 25.1\% of SSC A--B--A lineages contain exactly two triplets, with long tails extending to six triplets in Sigma and nine in SSC.

The median restore time (interval between $v_{i+1}$ and $v_{i+2}$) is 114.5 hours in Sigma and 159.6 hours in SSC.
In Sigma, 45.4\% of triplets are restored within 24 hours, compared to 22.3\% in SSC.
Fast restores (within hours) are consistent with immediate error correction, while longer intervals may reflect either delayed discovery of a problem or deliberate re-visitation of a prior structural form.

Manual inspection of representative cases in SSC suggests that these reverts often arise from alternative structural formulations of the same detection logic.
Compared to Sigma, SPL provides more expressive mechanisms for composing predicates and organizing logical structure, allowing equivalent conditions to be represented in many ways.
Additionally, repository-wide transformations such as field name standardization can temporarily shift rules into alternate structural forms before reverting to earlier representations.
We cannot determine the exact cause of each revert, but these patterns suggest that structural reversions reflect iterative refinement and re-expression of detection logic, as well as occasional correction of implementation ambiguity or defects, rather than simple rollback of mistakes.

\shortsection{Case study: AWS access key rule}
We illustrate how A--B--A reverts arise from multiple sources within a single SSC rule detecting AWS \keyword{CreateAccessKey} events where the acting user differs from the target user---a pattern indicative of credential misuse. The lineage contains nine A--B--A triplets, grouped into three episodes.

\shortersection{Field-name uncertainty (v13--v16, two triplets, July 2021)}
Four consecutive versions oscillate between two field names for the acting user---\keyword{userName} and \keyword{userIdentity.userName}---with each revert completing within hours. Both refer to the same CloudTrail attribute, suggesting uncertainty about canonical field choice rather than an intended change in detection logic.

\shortersection{Competing encodings of the same condition (v22--v35, six triplets, March--July 2022)}
Six triplets alternate between a direct inequality (\keyword{search userIdentity.userName!=requestParameters.userName}) and
an \keyword{eval}-based encoding (\keyword{eval match=if(match(...),1,0) | search match=0}).
The two forms express the same condition in different SPL idioms.
Notably, an initial \keyword{eval} attempt (v23) inverted the condition (\keyword{if(...,0,1)}), briefly flipping polarity before correction, showing how equivalent reformulations can introduce transient defects.

\shortersection{Transient typo (v46--v48, one triplet, July 2024)}
A single revision (v47) inserted a stray \keyword{change} token between \keyword{sourcetype=aws:cloudtrail} and \keyword{eventName=CreateAccessKey} that introduced an accidental free-text predicate which was removed in the next commit. This seems likely to be a typing error rather than intentional revision.
\\

Across all episodes, the detection intent---flagging mismatched actor and target users in \keyword{CreateAccessKey} events---remains unchanged. The structural churn reflects field ambiguity, equivalent encodings, and transient defects rather than substantive evolution.

%% file: 07_intent_analysis.tex
\section{Inferred Intent Behind Revisions}
\label{sec:intent_analysis}

Sections \ref{sec:temporal} and \ref{sec:structural} characterized when predicate-changing revisions occur and how they alter rule structure. Structural operations catalog that a predicate was added, removed, or reorganized, but not why. Revisions often reflect trade-offs between expanding coverage and reducing false positives. To study this semantic intent, we apply LLM inference to each adjacent version pair.

For each pair, we query GPT-5 with the two detection blocks and a structured prompt (see Appendix~\ref{appendix:llm-prompt} for details). The primary output is a \emph{rationale label}, one of four options: \op{coverage expansion} (CE), \op{false-positive reduction} (FPR), \op{mixed tradeoff} (MT, when a revision both broadens and narrows the matched event set), and \op{insufficient evidence} (IE). We also collect auxiliary signals: a coarse \emph{match-set direction} label (\op{broader}/\op{narrower}/\op{mixed}/\op{unclear}) and three Boolean flags for structural changes (\op{added}, \op{removed}, \op{modified}), which are not used in downstream analysis but support validation (\autoref{subsec:intent-validation}).

We restrict the analysis to pairs on which both \pgir and the LLM agree that predicate logic changed (agreement rate $\approx 99\%$), yielding 4{,}412 SSC pairs across 1{,}451 lineages and 7{,}958 Sigma pairs across 2{,}338 lineages.

We find that coverage expansion is the most common rationale in both corpora (41.3\% Sigma; 35.9\% SSC). At the lineage level, evolution is dominantly non-monotonic: more than half of multi-revision lineages reverse direction at least once across their history, and roughly a quarter oscillate repeatedly between coverage expansion and false-positive reduction (\autoref{subsec:intent-aggregate}). \autoref{subsec:intent-lineage-trajectories} illustrates three representative patterns through case studies.

\subsection{Validating the LLM Labels}
\label{subsec:intent-validation}

We assess the reliability of the rationale labels with two checks: an \emph{internal-consistency} check that the LLM's coarse direction label and rationale label tell the same story, and a \emph{cross-method-consistency} check that the LLM's structural claims align with \pgir\ operations derived independently.

\shortsection{Internal consistency}
\label{subsubsec:intent-internal}
Match-set direction and rationale label are emitted independently in the LLM's structured output: a \emph{broader} step should map to CE, \emph{narrower} to FPR, \emph{mixed} to MT, and \emph{unclear} to IE. We find this mapping is largely consistent but not perfect. Broader steps pair with CE in over 99\% of cases (3{,}289 of 3{,}310 Sigma; 1{,}586 of 1{,}597 SSC). Mixed steps map reliably to MT (95.4\% Sigma, 94.8\% SSC).  
Narrower steps map predominantly to FPR (85.6\% Sigma, 70.3\% SSC) but with some IE residual. Manual inspection of 10 such residual cases reveals two roughly equal sub-populations. In half, the matched set is genuinely unclear---the revision changes \emph{what} the rule observes rather than how strictly it observes it (\autoref{subsubsec:intent-ie-only}). In the other half, the matched set is observably narrower but the LLM still emits IE. These concentrate on revisions that narrow through \emph{structural enrichment} rather than direct predicate restriction. 
The downstream effect is a small under-count of FPR in favor of IE, particularly in SSC, where pipeline-driven narrowing is more common than in Sigma.

\shortsection{Cross-method consistency}
\label{subsubsec:intent-crossmethod}
The match-set direction and rationale label both originate from the LLM, so internal consistency cannot rule out a systematic LLM error in identifying that any change occurred. To measure confidence in the LLM labels, we also audit the LLM's three structural Booleans against the \pgir operations from \autoref{sec:structural}. 

\autoref{tab:intent-validation} reports the cross-method consistency check. For each adjacent version pair, we compare the three Boolean structural flags emitted by the LLM (\op{added}, \op{removed}, \op{modified}) against the structural operations recovered independently by the \pgir alignment.  The audit proceeds in two stages. First, for each pair, we check whether both methods agree that predicate-level change occurred. Second, on the agreed subset, we test whether the methods agree on the \emph{type} of change: \op{added} should align with \op{and+}/\op{or+}/\op{branch+}, \op{removed} with the corresponding removal operations, and \op{modified} with \op{val-update}; scope-changing operations (\op{move}, \op{flip}) can back any of the three.


\begin{table}[tb]
\caption{Validation of LLM structural claims. \rm Each corpus cell reports the number of rule steps labeled by the LLM with the corresponding output, followed in parentheses by the number and percentage of those steps supported by the \pgir analysis. 
}
\label{tab:intent-validation}
\centering
\small
\renewcommand{\arraystretch}{1.1}
\begin{threeparttable}
\begin{tabular}{lcc}
\toprule
\textbf{LLM output} & \textbf{Sigma} & \textbf{SSC} \\
\midrule
Total adjacent rule steps
& 39{,}477\tnote{a}
& 107{,}076 \\
\midrule
No predicate change
& \makecell[c]{31{,}503\\(31{,}225, 99.1\%)}
& \makecell[c]{101{,}694\\(101{,}440, 99.8\%)} \\[2.5ex]
Predicate change
& \makecell[c]{7{,}974\\(7{,}956, 99.8\%)}
& \makecell[c]{5{,}382\\(4{,}412, 82.0\%)} \\[2.5ex]
\midrule
Addition flag
& \makecell[c]{3{,}615\\(3{,}168, 87.6\%)}
& \makecell[c]{2{,}008\\(1{,}950, 97.1\%)} \\[2.5ex]
Removal flag
& \makecell[c]{2{,}356\\(2{,}076, 88.1\%)}
& \makecell[c]{1{,}449\\(1{,}393, 96.1\%)} \\[2.5ex]
Modification flag
& \makecell[c]{5{,}127\\(3{,}812, 74.4\%)}
& \makecell[c]{2{,}617\\(1{,}824, 69.7\%)} \\
\bottomrule
\end{tabular}
\begin{tablenotes}[flushleft]
\footnotesize
\item[a] We exclude 3 Sigma pairs because the corresponding prompts exceed the model context window.
\end{tablenotes}
\end{threeparttable}
\end{table}

Pair-level agreement is extremely high for no predicate change (99.1\% in Sigma; 99.8\% in SSC). Agreement for whether there is \emph{any} change in the predicate is also high for Sigma (99.8\%), but lower in SSC (82.0\%). Manual inspection indicates that many SSC disagreements arise because the LLM sometimes treats changes to data-transformation or presentation pipelines as predicate-level changes, whereas \pgir only counts changes that affect filtering predicates. Within the subset where both LLM and \pgir agree that at least one predicate-level change occurred, the per-claim support rate from \pgir is high for additions and removals but lower for modifications (74.4\% Sigma, 69.7\% SSC). Manual inspection indicates that cases where LLM outputs a modification flag but \pgir doesn't show aligned operations concentrate on a single mechanism: revisions that change values \emph{inside} an existing predicate---for example, adding a filename or command pattern to an existing \keyword{IN} or \keyword{contains} expression. The LLM treats such edits as new predicates entering the rule, while \pgir records them as \op{val-update} because no new conjunct, disjunct, or branch is introduced. The same mechanism accounts for the elevated Sigma addition mismatch rate, since Sigma rules frequently evolve by extending or contracting long value lists. These mismatches reflect a representational ambiguity between semantic and structural revision rather than an LLM annotation failure, and they do not affect the rationale label, which is the only LLM output used downstream.

\shortsection{Model selection} 
We sampled 20 pairs (10 from each repository) from \pgir-flagged changed pairs, and experimented with GPT-4o-mini, GPT-4o, and GPT-5 on separating detection logic and identify changes across pairs. Only GPT-5 was able to reliably decouple filtering pipeline stages from downstream formatting stages. All of the results reported are using GPT-5 as the LLM.

\subsection{Aggregate Intent Distribution}
\label{subsec:intent-aggregate}

Having established that the labels output by the LLM are fairly reliable, we next examine what kinds of operational intent they reveal. \autoref{tab:intent-summary} summarizes the results at two levels: the pair-level distribution of \keyword{rationale label} and the lineage-level taxonomy of aggregate intent trajectories.

\shortsection{Revisions} 
Coverage expansion is the most common rationale in both corpora, accounting for 41.3\% of Sigma pairs (3{,}289 of 7{,}958) and 35.9\% of SSC pairs (1{,}586 of 4{,}412). False-positive reduction is the next most frequent in Sigma (30.2\%, 2{,}401 of 7{,}958) but is markedly less common in SSC (18.0\%). The IE rate, conversely, is much higher in SSC (34.6\%) than in Sigma (16.5\%). Mixed tradeoffs are comparable across corpora (12.0\% Sigma, 11.4\% SSC).
The IE--FPR asymmetry between corpora partially reflects the labeling artifact identified in \autoref{subsubsec:intent-internal}; the SSC pair-level FPR rate should be read as a lower bound since many revisions classified as IE are likely FPRs.

\shortsection{Rule lineages}
To characterize lineage-level trajectories, we first distinguish two layers at which a revision sequence can mix broadening and narrowing intent: \emph{within-revision} mixing is captured by the MT label where a single edit both broadens and narrows the matched event set; \emph{across-revision} mixing instead arises when distinct revisions in a lineage push in opposing directions, even if no individual revision is itself MT. We use Coupled for lineages dominated by within-revision MT and Alternating for lineages exhibiting across-revision direction changes.


\autoref{tab:intent-summary} partitions lineages into three disjoint cohorts: IE-only (insufficient evidence for every revision), Singleton (exactly one labeled (non-IE) revision), and Multi-revision (at least two labeled revisions). Within the IE-only cohort, single-revision lineages dominate:  77.4\% of Sigma IE-only lineages (192 of 248)  and
68.9\% of SSC (146 of 212) contain exactly one revision, and fewer than 3\% in either corpus contain more than four. 
The small tail of longer IE-only lineages (up to 14 revisions in SSC, 5 in Sigma) reflects sustained representational ambiguity (\autoref{subsubsec:intent-ie-only}). 

Trajectory analysis is meaningful only within the multi-revision cohort (676 SSC and 1{,}245 Sigma lineages) since a trajectory requires at least two directional data points.
For multi-revision lineages we apply a priority-ordered classification, assigning each lineage a single label based on the first matching rule. A lineage is Coupled if at least 50\% of its non-IE revisions are MT. The priority of this rule means a lineage is labeled as Coupled regardless of whether its non-MT revisions also alternate, since its dominant mixed character is within-revision rather than across-revision. Lineages not meeting this criterion are classified by their directional subsequence as CE-only if every directional revision is CE, FPR-only if every directional revision is FPR, and Alternating if both appear. Alternating lineages are further sub-classified by the number of direction changes $\tau$ in the directional subsequence, either Oscillating ($\tau \geq 2$), where the direction flips back at least once (e.g., CE$\rightarrow$FPR$\rightarrow$CE) or Phased ($\tau = 1$), where a single regime change occurs with no return (e.g., CE$\rightarrow$CE$\rightarrow$FPR$\rightarrow$FPR).

The distribution, reported in \autoref{tab:intent-summary}, shows that alternating trajectories dominate, comprising 56\% of multi-revision lineages in both rulesets.  
Across all rule lineages, 29.8\% of Sigma lineages and 26.2\% of SSC lineages alternate between coverage expansion and false-positive reduction at some point in their history. Pure-direction trajectories, in which every revision pushes the same way, are a clear minority (CE-only 25.1\% Sigma, 23.1\% SSC; FPR-only 6.9\% Sigma, 3.1\% SSC). 
Secondly, \emph{repeated reversal is common}: oscillating lineages account for 27.6\% of multi-revision Sigma lineages (344) and 23.5\% of SSC (159). Coupled lineages, in which most revisions are themselves MT, are less common but qualitatively distinct (12.0\% Sigma, 17.6\% SSC). The case studies in \autoref{subsec:intent-lineage-trajectories} provide examples of one Oscillating, one Coupled, and one IE-only lineage. 

\begin{table}[tb]
\centering
\caption{Intent categories and trajectory patterns.}
\label{tab:intent-summary}
\small
\setlength{\tabcolsep}{6pt}
\renewcommand{\arraystretch}{0.96}
\begin{tabularx}{0.9\columnwidth}{@{}p{1em}p{14em}rr}
\toprule
 & \textbf{Category/Pattern} &
 \multicolumn{1}{c}{\textbf{Sigma}} &
 \multicolumn{1}{c}{\textbf{SSC}} \\
\midrule
\multirow{5}{*}{\rotatebox[origin=c]{90}{Revisions}}
& Number of revision pairs & 7{,}958 & 4{,}412 \\
& Coverage expansion (CE)        & 41.3\% & 35.9\% \\
& False-positive reduction (FPR) & 30.2\% & 18.0\% \\
& Mixed tradeoff (MT)            & 12.0\% & 11.4\% \\
& Insufficient evidence (IE)     & 16.5\% & 34.6\% \\
\midrule
\multirow{11}{*}{\rotatebox[origin=c]{90}{Lineages}}
& Number of rule lineages & 2{,}338 & 1{,}451 \\
& \multicolumn{3}{@{}c@{}}{\textit{Cohorts (\% of all)}} \\
& \quad IE-only        & 10.6\% & 14.6\% \\
& \quad Singleton      & 36.1\% & 38.8\% \\
& \quad Multi-revision & 53.3\% & 46.6\% \\
\cmidrule(l){2-4}
& Multi-revision lineages & 1{,}245 & 676 \\

& \multicolumn{3}{@{}c@{}}{\textit{Trajectory (\% of multi-revision lineages)}} \\
& \quad Coupled ($\geq 50\%$ MT)        & 12.0\% & 17.6\%  \\
& \quad CE-only                         & 25.1\% & 23.1\%  \\
& \quad FPR-only                        & 6.9\%  & 3.1\%  \\
& \quad Alternating                     & 56.0\% & 56.2\% \\
& \quad\quad Oscillating ($\tau \geq 2$) & 27.6\% & 23.5\%  \\
& \quad\quad Phased ($\tau = 1$)         & 28.4\% & 32.7\%  \\
\bottomrule
\end{tabularx}
\end{table}

\shortsection{Timing}
We measure transition gaps as the elapsed time between two consecutive directional revisions (CE or FPR) within a lineage, ignoring intervening IE and MT revisions. 
This restricts the measurement to revisions whose direction can be determined from the rule text. We note that IE is a heterogeneous category covering revisions whose direction is genuinely undefined as well as revisions with a direction that is not determined by the LLM from the rule text (\autoref{subsubsec:intent-ie-only}). Thus, the gaps reported here measure cadence between direction-attributable revisions specifically, not the total cadence of operationally meaningful change.

Sustained coverage maintenance proceeds on a markedly slower cadence than sustained false-positive suppression. As shown in \autoref{tab:intent-timing}, the median CE$\rightarrow$CE gap is 125 days for Sigma and 187 days for SSC, whereas FPR$\rightarrow$FPR gaps span only 32 days in Sigma and 110 days in SSC. This is consistent with FPR being a reactive maintenance mode in which false positives surface in deployment within days to weeks and trigger immediate patches, whereas CE follows a slower planning cadence as new threats are discovered. 

Oscillating lineages exhibit two temporal signatures. First, the initial directional reversal occurs early (median 144 days in Sigma, 51 days in SSC), indicating that the underlying tension emerges within the first year rather than after prolonged deployment. Second, most do not settle within the observation window: 58.4\% of Sigma and 74.2\% of SSC oscillators are still flipping at the snapshot date. Together, these findings are consistent with the case study (\autoref{subsubsec:intent-oscillating}) and suggest that the tension is rarely resolved through repeated revisions.


\begin{table}[tb]
\centering
\caption{Transition gaps and oscillator timing. \rm Transition gaps are the median elapsed days between two consecutive directional revisions (CE/FPR) (intervening IE/MT revisions are skipped).}
\label{tab:intent-timing}
\small
\setlength{\tabcolsep}{3pt}
\renewcommand{\arraystretch}{0.95}
\begin{tabularx}{\columnwidth}{@{}Xcc@{}}
\toprule
& \multicolumn{1}{c}{\textbf{Sigma}} & 
\multicolumn{1}{c}{\textbf{SSC}} \\
\midrule
\textit{Transition gap (median days)} & & \\
\quad CE $\rightarrow$ CE; FPR $\rightarrow$ FPR   & 125; 32 & 187; 110 \\
\quad CE $\rightarrow$ FPR; FPR $\rightarrow$ CE   & $\,\,\,$44; 74  & $\,\,\,$21; $\,\,\,$68 \\
\midrule
\textit{Oscillating lineages} & & \\
\quad Time to first flip (median) & 144 days & 51 days \\
\quad Still flipping at snapshot       & 58.4\% (201/344) & 74.2\% (118/159) \\
\bottomrule
\end{tabularx}
\end{table}

\subsection{Case Studies}
\label{subsec:intent-lineage-trajectories}

To make these trajectory types concrete, we describe three representative cases: an oscillating lineage (from SSC), a Coupled lineage (from Sigma), and an IE-only lineage (from Sigma). Each illustrates a distinct mechanism behind non-monotonic rule evolution.

\shortsection{Oscillation: strategy conflict}
\label{subsubsec:intent-oscillating}

Oscillating lineages are not merely alternating in aggregate, they repeatedly transition between broader and narrower revisions as authors revisit the same unresolved design choice. The SSC example with the most direction changes is \keyword{detect\_new\_local\_admin\_account}, which detects creation of a new local administrator account by correlating two Windows security events on the same host: \keyword{EventCode=4720} (user account created) and \keyword{EventCode=4732} (member added to local Administrators). Across its history, the rule alternates between two post-search pipelines that encode different answers to the same question: {\em should the detector group the two events by temporal proximity, or require both event types?} 

The two formulations have substantially different semantics. The \keyword{transaction}-based form groups events within a time window and emits a row even if only one event type is present, making it broader. The \keyword{stats}+\keyword{evCount=2} form aggregates by user--destination pair and explicitly enforces co-occurrence, making it narrower. The oscillation is most clearly visible in three consecutive late revisions (the search prefix \keyword{\textquotesingle wineventlog\_security\textquotesingle\ EventCode=4720 OR (EventCode=4732 Group\_Name=Administrators)} is shared across all versions and elided below):

\begin{lstlisting}[
  basicstyle=\sffamily\small,
  breaklines=true, breakatwhitespace=false, columns=flexible,
  escapeinside={(*}{*)},
  xleftmargin=0pt,
  breakindent=2em,
  aboveskip=4pt, belowskip=4pt
]
(*\textrm{\normalsize {\bf v61 (CE)}: transaction-based grouping}*) 
... | transaction member_id connected=false maxspan=180m | rename member_id as user | stats count min(_time) as firstTime max(_time) as lastTime by user dest
(*\textrm{\normalsize {\bf v62 (FPR)}: explicit co-occurrence gate}*)
... | stats dc(EventCode) as evCount min(_time) as _time range(_time) as duration values(src_user) as src_user by user dest | where evCount=2 AND duration<7200 | fields - evCount, duration
(*\textrm{\normalsize {\bf v63 (CE)}: reverted to transaction (same as v61)}*)
\end{lstlisting}

About 13 months later, v87$\rightarrow$v88 attempts a hybrid that retains \keyword{transaction} grouping but adds an explicit distinct \keyword{EventCode} constraint (\keyword{dc(EventCode)>1}) on the grouped output---suggesting an eventual recognition that neither pure formulation was satisfactory. 

The rule maintainers are not drifting aimlessly; they are repeatedly switching between competing implementations of the same detection goal. The oscillation arguably reflects a design problem: the broad and narrow formulations encode different operational priorities, and a more durable resolution might split the two into separate rules feeding distinct triage queues. Absent that refactor, and absent deployment feedback on which formulation yields the better precision--coverage tradeoff, rule maintainers return to the same unresolved choice across years.

\shortsection{Coupled: expansion and exclusion growth} Coupled lineages reflect a different mechanism: revisions whose broadening and narrowing effects are structurally coupled within the same step. A clear example is the Sigma rule \keyword{proc\_creation\_win\_renamed\_binary\_highly\_relevant}, which detects renamed living-off-the-land binaries (LOLBins) by matching embedded identity fields (\keyword{OriginalFileName}, later also \keyword{Description} and \keyword{Product}) against a target list of sensitive binary names, while excluding executions whose image path matches a legitimate binary location. The lineage is overwhelmingly Coupled: 6 out of 7 of its non-IE revisions carry the MT label. 

With this approach, every newly targeted binary identity also creates a new benign execution surface that must be excluded. The first clear MT step (v10$\rightarrow$v11) adds \keyword{pwsh.dll} to the positive identity list while simultaneously adding \keyword{*\textbackslash pwsh.exe} to the exclusion list:

\begin{lstlisting}[
  basicstyle=\sffamily\small,
  breaklines=true, breakatwhitespace=false, columns=flexible,
  escapeinside={(*}{*)},
  xleftmargin=0pt,
  breakindent=2em,
  aboveskip=4pt, belowskip=4pt
]
(*\textrm{\normalsize {\bf v11 (MT)}: pwsh added to both inclusion and exclusion}*)
OriginalFileName IN ("powershell.exe", "pwsh.dll", "psexec.exe", ...) NOT (Image IN ("*\\powershell.exe", "*\\pwsh.exe", "*\\psexec.exe", ...))
\end{lstlisting}

The same paired-addition mechanism recurs across the lineage---v15$\rightarrow$v16 adds \keyword{psexesvc.exe} to the target set alongside \keyword{*\textbackslash PSEXESVC.exe} in the exclusion set, with later revisions repeating the pattern for \keyword{reg.exe}, \keyword{msxsl.exe}, and others. Unlike the oscillating SSC case, this lineage does not alternate between competing goals; its bidirectionality is intrinsic to the rule's authoring logic. Expanding the suspicious identity surface necessarily requires growing the exclusion surface in parallel, so the repeated MT labels reflect structural coupling rather than indecision.

\shortsection{Insufficient evidence: multiple mechanisms behind unrecoverable direction}
\label{subsubsec:intent-ie-only}

\renewcommand{\_}{\discretionary{\char`\_}{}{\char`\_}}

IE-only lineages highlight a third source of non-monotonicity: revisions that are clearly visible syntactically but do not expose a recoverable directional effect from rule text alone. The Sigma rule \keyword{win\_defender\_config\_change\_exclusion\_added} illustrates two distinct sub-mechanisms within a single lineage. The rule flags additions to the Windows Defender exclusion list, but its successive revisions express that intent through different telemetry sources and different field names:

\begin{lstlisting}[
  basicstyle=\sffamily\small,
  breaklines=true, breakatwhitespace=false, columns=flexible,
  escapeinside={(*}{*)},
  xleftmargin=0pt,
  breakindent=2.5em,
  aboveskip=4pt, belowskip=4pt
]
(*\textrm{\normalsize \bf v3:}*)  EventID=13 AND TargetObject="*\\Microsoft\\Windows Defender\\Exclusions*"
(*\textrm{\normalsize \bf v4:}*)  EventID=5007 AND "New Value"="*\\Microsoft\\Windows Defender\\Exclusions*"
(*\textrm{\normalsize \bf v5:}*)  EventID=5007 AND New_Value="*\\Microsoft\\Windows Defender\\Exclusions*"
(*\textrm{\normalsize \bf v10:}*) EventID=5007 AND NewValue="*\\Microsoft\\Windows Defender\\Exclusions*"
\end{lstlisting}
\noindent
The v3$\rightarrow$v4 step is a \emph{telemetry source migration}. The rule switches from Sysmon-recorded registry events (\keyword{EventID=13}) to Windows Defender's native configuration-change events (\keyword{EventID=5007}). These are disjoint event streams with different population, semantics, and schema; whether the rewrite broadens or narrows the matched event set depends on which source is actually populated in deployment. The subsequent steps (v4$\rightarrow$v5, v5$\rightarrow$v10) are \emph{field renaming} within the same source, where the same wildcard target moves between different conventions for naming the registry-value field. 
A third mechanism, \emph{detection target redirection}, appears in the SSC rule \keyword{cloud\_excessive\_provisioning\_activities}, which detects anomalous bursts of cloud instance activity. Between v2 and v3 the rule pivots from anomalous \emph{provisioning} to anomalous \emph{destruction} on otherwise unchanged scaffolding:

\begin{lstlisting}[
  basicstyle=\sffamily\small,
  breaklines=true, breakatwhitespace=false, columns=flexible,
  escapeinside={(*}{*)},
  xleftmargin=0pt,
  breakindent=2.5em,
  aboveskip=4pt, belowskip=4pt
]
(*\textrm{\bf \normalsize v2:}*) All_Changes.action=created OR All_Changes.action=started AND All_Changes.status=success AND All_Changes.object_category=instance
(*\textrm{\bf \normalsize v3:}*) All_Changes.action=deleted AND All_Changes.status=success AND All_Changes.object_category=instance
\end{lstlisting}

The matched event sets of v2 and v3 are disjoint: the provisioning events and destruction events do not overlap. The revision is operationally meaningful---v3 detects a different attacker behavior (e.g., ransomware-style cleanup or anti-forensic cleanup) than v2 (e.g., cryptojacking-style spinup)---but neither version's matched set contains the other, so no broadening or narrowing relation exists.

In all three mechanisms, the revision changes \emph{what} the rule observes rather than \emph{how strictly} it observes it, so there is no clear direction to the change.

\shortsection{Discussion}
Together, these three case studies surface three distinct mechanisms behind non-monotonic rule evolution. Oscillating lineages expose unresolved \emph{precision--coverage tension}: competing implementations repeatedly replace one another because the better operational choice depends on deployment feedback that varies over time. Coupled rules expose a \emph{stable maintenance regime} in which coverage expansion and exclusion growth are intrinsically coupled by the detection strategy itself, so paired broadening and narrowing is a structural necessity rather than reflecting indecision. IE-only rules expose a \emph{representational limit}: some revisions are visible and likely operationally meaningful, yet their directional effect cannot be inferred from rule text alone without external knowledge of schemas, parsers, or telemetry populations. Taken together, these mechanisms show that rule maintenance in public repositories is multi-objective and path-dependent. Rule lineages do not typically exhibit a gradual drift toward a single target and a significant number of rules carry unresolved design tensions that persist across years of revision. Neither the structural analysis nor the intent analysis alone could have produced this characterization: it requires reading structural signals through the lens of inferred intent.

%% file: 08_discussion.tex
\section{Conclusion}
\label{sec:discussion}

Our longitudinal analysis of log-based detection rules 
reflects ongoing operational trade-offs rather than steady convergence. Detection rule evolution is sparse per commit but cumulative over rule lifetimes, structurally coordinated rather than incremental, and frequently non-monotonic.
Our findings raise sharper questions than any static view of detection logic permits: 
\begin{itemize} \renewcommand{\labelitemi}{--}
\item why do many rules revisit the same precision--coverage choice across years? 
\item when does paired expansion-and-exclusion reflect intentional coupling between coverage and false-positive control versus unresolved indecision? 
\item how much of detection engineering remains invisible to text-level analysis, locked behind schemas, parsers, and telemetry data that only deployment makes legible? 
\end{itemize}

Answering these questions requires infrastructure the field currently lacks: tooling that distinguishes refactorings from genuine semantic changes and compares rule versions on detection logic rather than surface forms, and AI-assisted authoring and review systems that operate over canonicalized representations rather than raw rule text.
Closing the loop with deployment feedback that links revision histories to false-positive rates, alert volumes, and analyst dispositions in real SOCs is an important step toward better processes for devising and deploying security rules.

%% file: appendix.tex
\section{\pgir\ Example}\label{appendix:pgir-example}

The \pgir\ extracted from the SPL example in \autoref{subsec:rules-bg} is shown below. The rule contains two filtering predicates, both from the root search stage. Each leaf is shown as a typed atomic predicate in the form \keyword{PRED(field, operator, value)}.

\begin{lstlisting}[basicstyle=\sffamily\small,
  breaklines=true, columns=flexible,
  aboveskip=12pt, belowskip=4pt]
Rule: detections/endpoint/linux_auditd_sudo_or_su_execution.yml
Repo: SSC
Version: 31
Commit: 11c909f725435e69e87cc7fde4558fb366432dc1
Predicate count: 2

Predicate graph:
EXPR(op=AND)
  PRED(field=sourcetype,
       operator=EQ,
       value=STRING("auditd"))
  PRED(field=proctitle,
       operator=IN,
       value=[WILDCARD("*sudo *"),
              WILDCARD("*su *")])
\end{lstlisting}

\section{Algorithm for Phase Matching}
\label{appendix:align-algo}

\autoref{alg:align} provides full pseudocode for the four-phase
alignment procedure described in \autoref{subsec:alignment}. The algorithm
incrementally extends a partial mapping $\Phi$ between the canonicalized
predicate trees of two rule versions: Phase~1 anchors globally unique
predicate leaves, Phase~2 propagates this evidence to operator nodes
bottom-up, Phase~3 resolves locally unique duplicates within already-matched
operator scopes, and Phase~4 admits conservative fuzzy matches before
re-running Phase~2 to recover any operator correspondences newly supported
by fuzzy evidence. The auxiliary \keyword{MatchOperators} procedure
encapsulates the operator-matching logic shared between the two Phase~2
invocations and takes an \keyword{evidenceMode} flag that controls whether
only exact anchors or all current matches are used as evidence.

\begin{algorithm*}[p]
\caption{PGIR Predicate Tree Alignment}
\label{alg:align}
\small
\DontPrintSemicolon
\SetCommentSty{textrm}
\SetKwComment{AlgComment}{$\triangleright$ }{}
\SetKw{KwAnd}{and}
\SetKw{KwOr}{or}
\SetKw{KwNot}{not}
\SetKwProg{Proc}{procedure}{}{end}

\KwIn{PGIR rule versions $A$, $B$}
\KwOut{Partial mapping $\Phi: \mathrm{nodes}(T_A) \rightharpoonup \mathrm{nodes}(T_B)$, unmatched nodes}

\vspace{3pt}
$T_A \leftarrow$ \keyword{Canonicalize}$(A)$;\quad
$T_B \leftarrow$ \keyword{Canonicalize}$(B)$\;
$\Phi \leftarrow \emptyset$;\quad $\mathit{UsedOpB} \leftarrow \emptyset$;\quad $\mathit{UsedPredB} \leftarrow \emptyset$\;

\vspace{3pt}
\AlgComment*[l]{Phase 1: global exact predicate anchors}
build \keyword{ExactKey} index for predicate leaves in $T_A$ and $T_B$\;
\AlgComment*[l]{\keyword{ExactKey}$(p) = (\mathit{field},\ \mathit{cmp},\ \mathit{norm(value)},\ \mathit{polarity})$; $\mathit{cmp} \in \{\textsc{eq}, \textsc{in}, \textsc{regex}, \ldots\}$}
\ForEach{key $k$ that is globally 1-to-1 between $T_A$ and $T_B$}{
    let $p$ be the unique $A$-leaf with \keyword{ExactKey}$(p) = k$\;
    let $q$ be the unique $B$-leaf with \keyword{ExactKey}$(q) = k$\;
    $\Phi[p] \leftarrow q$;\quad
    $\mathit{UsedPredB} \leftarrow \mathit{UsedPredB} \cup \{q\}$;\quad
    mark $p$, $q$ as anchors\;
}

\vspace{3pt}
\AlgComment*[l]{Phase 2: operator scope matching (bottom-up), using exact-anchor evidence only}
\keyword{MatchOperators}$(T_A, T_B, \Phi, \mathit{UsedOpB},\ \mathit{evidenceMode} = \textsc{exact})$\;

\vspace{3pt}
\AlgComment*[l]{Phase 3: exact duplicate completion within matched operator scopes}
\ForEach{matched operator pair $(o_A \mapsto o_B) \in \Phi$}{
    let $k$ appear $m \geq 1$ times among unmatched leaves under $o_A$ \KwAnd $m$ times under $o_B$\;
    let $i_1,\ldots,i_m$ and $j_1,\ldots,j_m$ be those leaves in sorted order\;
    \lForEach{$t = 1,\ldots,m$}{add $(i_t \mapsto j_t)$ to $\Phi$ if ancestry-consistent; update $\mathit{UsedPredB}$}
}

\vspace{3pt}
\AlgComment*[l]{Phase 4: conservative fuzzy predicate matching}
\ForEach{unmatched predicate leaf $p \in T_A$}{
    $C \leftarrow$ \keyword{CoarseCandidates}$(p,\ T_B \setminus \mathit{UsedPredB})$\;
    \AlgComment*[r]{gated by polarity, value type, \keyword{CmpClass}; capped at $K$ candidates}
    split $C$ into same-field $C_s$ and cross-field $C_x$\;
    $q^* \leftarrow$ \keyword{BestMatch}$(p, C_s, \Phi)$ if exists, else \keyword{BestMatch}$(p, C_x, \Phi)$\;
    \lIf{$q^*$ exists}{$\Phi[p] \leftarrow q^*$;\quad $\mathit{UsedPredB} \leftarrow \mathit{UsedPredB} \cup \{q^*\}$}
}

\vspace{3pt}
\AlgComment*[l]{Phase 2 again: operator completion using all mapped predicates as evidence}
\keyword{MatchOperators}$(T_A, T_B, \Phi, \mathit{UsedOpB},\ \mathit{evidenceMode} = \textsc{all})$\;

\vspace{3pt}
\KwRet $\Phi$ and unmatched nodes in $T_A$, $T_B$\;

\vspace{6pt}
\Proc{\normalfont\keyword{MatchOperators}$(T_A, T_B, \Phi, \mathit{UsedOpB}, \mathit{evidenceMode})$}{
    \ForEach{operator $o_A \in T_A$ in increasing height order}{
        \lIf{$o_A \in \mathrm{dom}(\Phi)$}{\textbf{skip}}\;
        \ForEach{candidate $o_B \in T_B$: $\mathrm{label}(o_B) = \mathrm{label}(o_A)$, $o_B \notin \mathit{UsedOpB}$, ancestry-consistent}{
            $E_A \leftarrow$ evidence predicates under $o_A$ per $\mathit{evidenceMode}$\;
            $E_B \leftarrow$ evidence predicates under $o_B$ per $\mathit{evidenceMode}$\;
            \lIf{$|E_A| < \mathit{minAnchors}$ \KwOr $|E_B| < \mathit{minAnchors}$}{\textbf{skip}}\;
            $\mathit{overlap} \leftarrow |\{\,p \in E_A : \Phi[p]\ \text{is a descendant of}\ o_B\,\}|$\;
            $\mathit{support} \leftarrow \mathit{overlap}\ /\ \min(|E_A|,\ |E_B|)$\;
            $\mathit{cov}_A \leftarrow \mathit{overlap}\ /\ |E_A|$;\quad $\mathit{cov}_B \leftarrow \mathit{overlap}\ /\ |E_B|$\;
            \lIf{$\mathit{support} < \theta_{\mathit{sup}}$ \KwOr $\mathit{cov}_A < \theta_{\mathit{cov}}$ \KwOr $\mathit{cov}_B < \theta_{\mathit{cov}}$}{\textbf{skip}}\;
            score $o_B$ by $(\mathit{support},\ {-}|h(o_A) - h(o_B)|)$ (tiebreak: prefer symmetric height)\;
        }
        \lIf{best-scoring $o_B^*$ exists}{$\Phi[o_A] \leftarrow o_B^*$;\quad $\mathit{UsedOpB} \leftarrow \mathit{UsedOpB} \cup \{o_B^*\}$}
    }
}
\end{algorithm*}

\section{Mimikatz Rule History}
\label{appendix:mimikatz-history}

This appendix expands on the cohort-level outlier flagged in
\autoref{subsubsec:cohort}, where the 2016-Q4 Sigma cohort accumulates an order of magnitude more lifetime edit magnitude per rule than any other cohort. The bulk of that mass comes from a single rule,
\keyword{win\_alert\_mimikatz\_keywords.yml}, whose history we trace
below. The trajectory illustrates how a single late-life expand-then-contract episode in one rule can dominate cohort-level statistics for a small cohort, and why the magnitude figure should not be read as evidence of unusually heavy maintenance across the cohort as a whole.

The rule was introduced on 2016-12-27 and spans 41 observed versions, 15 predicate-changing transitions, and a total predicate edit magnitude of $\sum d_{\text{step}} = 519.2$. Most of this mass is not accumulated
gradually; it comes from a short expand-then-contract episode in late
2021 and early 2022, summarized below.

\paragraph{Early incremental maintenance.}
For most of its history, the rule remained a compact keyword detector for
Mimikatz-related command-line or log-message strings. Early edits changed
the matching context or modestly extended the keyword list: v4$\rightarrow$v5
added Sysmon coverage and \keyword{mimidrv.sys}; v9$\rightarrow$v10 removed
the explicit \keyword{EventLog} selection; v13$\rightarrow$v14 converted
terms into wildcarded matches; and v16$\rightarrow$v17 bound the list to
the \keyword{Message} field. A later normalization rewrote the same
matching behavior using \keyword{Message|contains}. These edits account for
visible maintenance activity but do not explain the lineage's outlier
status.

\paragraph{False-positive filtering.}
The first larger structural change occurs shortly before the outlier
episode. At v25$\rightarrow$v26, the rule keeps the same keyword list but
adds a filter excluding Sysmon Event ID~15, changing the condition from a
pure keyword match to \keyword{keywords and not filter}:

\begin{lstlisting}[
  basicstyle=\sffamily\small,
  breaklines=true, breakatwhitespace=false, columns=flexible,
  escapeinside={(*}{*)},
  xleftmargin=0pt,
  breakindent=2em,
  aboveskip=4pt, belowskip=4pt
]
(*\textrm{\normalsize \bf v25:}*) "\\mimikatz" OR "mimikatz.exe" OR "\\mimilib.dll" OR ... OR "gentilkiwi.com" OR "Kiwi Legit Printer"
(*\textrm{\normalsize \bf v26:}*) "\\mimikatz" OR "mimikatz.exe" OR "\\mimilib.dll" OR ... OR "gentilkiwi.com" OR "Kiwi Legit Printer" NOT EventID=15
\end{lstlisting}

This edit has $d_{\text{pred}}=39.0$ and explicitly introduces a
false-positive reduction mechanism, but the rule is still organized around
a short keyword list.

\shortsection{Large expansion}
The main outlier jump occurs at v26$\rightarrow$v27 on 2021-12-20, with
$d_{\text{pred}}=195.0$. This revision preserves the same high-level rule
skeleton but expands the positive keyword set from a short list of
Mimikatz indicators into a broad catalogue of Mimikatz subcommands and
adjacent tool strings:

\begin{lstlisting}[
  basicstyle=\sffamily\small,
  breaklines=true, breakatwhitespace=false, columns=flexible,
  escapeinside={(*}{*)},
  xleftmargin=0pt,
  breakindent=2em,
  aboveskip=4pt, belowskip=4pt
]
(*\textrm{\normalsize \bf v26:}*) "\\mimikatz" OR "mimikatz.exe" OR "\\mimilib.dll" OR "privilege::debug" OR "sekurlsa::logonpasswords" OR "lsadump::sam" OR "gentilkiwi.com" OR "Kiwi Legit Printer" NOT EventID=15
(*\textrm{\normalsize \bf v27:}*) "\\mimikatz" OR "mimikatz.exe" OR "\\mimilib.dll" OR "sekurlsa::logonpasswords" OR "crypto::capi" OR "crypto::certificates" OR "dpapi::blob" OR "dpapi::chrome" OR "kerberos::golden" OR "lsadump::dcsync" OR "misc::skeleton" OR "net::user" OR "privilege::debug" OR "process::list" OR "rpc::server" OR "service::start" OR "sid::add" OR "standard::base64" OR "token::elevate" OR "ts::sessions" OR "vault::list" ... NOT EventID=15
\end{lstlisting}

The expansion adds roughly 181 predicates, covering large command families
such as \keyword{crypto::}, \keyword{dpapi::}, \keyword{kerberos::},
\keyword{lsadump::}, \keyword{misc::}, \keyword{net::},
\keyword{privilege::}, \keyword{process::}, \keyword{rpc::},
\keyword{service::}, \keyword{sid::}, \keyword{standard::},
\keyword{token::}, and \keyword{ts::}. Several small cleanup edits follow
on the same date, removing duplicate or excess entries, but the rule
remains in a highly expanded state.

\shortsection{Rapid contraction}
The expansion is sharply rolled back two weeks later. At v30$\rightarrow$v31
on 2022-01-05, the rule removes approximately 153 predicates
($d_{\text{pred}}=166.6$). The commit message cites the
\emph{``massive performance impact of keyword-based rule''}, and the new
version replaces many specific subcommands with a much smaller set of
broader tokens, including prefix-style entries such as
\keyword{lsadump::} and \keyword{sekurlsa::}.

\begin{lstlisting}[
  basicstyle=\sffamily\small,
  breaklines=true, breakatwhitespace=false, columns=flexible,
  escapeinside={(*}{*)},
  xleftmargin=0pt,
  breakindent=2em,
  aboveskip=4pt, belowskip=4pt
]
(*\textrm{\normalsize \bf v30:}*) "crypto::capi" OR "crypto::certificates" OR "crypto::certtohw" OR "dpapi::blob" OR "dpapi::chrome" OR "kerberos::ask" OR "kerberos::golden" OR "lsadump::backupkeys" OR "lsadump::dcsync" OR "lsadump::sam" OR "misc::printnightmare" OR "net::user" OR "privilege::debug" OR "sekurlsa::logonpasswords" OR "token::elevate" OR "vault::list" ... NOT EventID=15
(*\textrm{\normalsize \bf v31:}*) "crypto::certificates" OR "crypto::tpminfo" OR "dpapi::masterkey" OR "kerberos::golden" OR "kerberos::ptc" OR "kerberos::ptt" OR "kerberos::tgt" OR "lsadump::" OR "mimidrv.sys" OR "\\mimilib.dll" OR "misc::printnightmare" OR "misc::shadowcopies" OR "privilege::backup" OR "privilege::debug" OR "privilege::driver" OR "sekurlsa::" NOT EventID=15
\end{lstlisting}

A second edit on the same date, v31$\rightarrow$v32
($d_{\text{pred}}=20.0$), reduces the set further, leaving a compact
performance-conscious approximation rather than an enumerated command
catalogue.

\shortsection{Interpretation}
This history shows why the 2016-Q4 cohort should be interpreted as a
small-cohort outlier rather than as evidence that early Sigma rules
generally accumulated unusually large predicate changes. The cohort-level mass is largely produced by one operational episode in one rule: an attempt to broaden a keyword detector into an enumerated Mimikatz command catalogue, followed almost immediately by pruning motivated by performance cost.

\section{LLM Prompt for Intent Inference}
\label{appendix:llm-prompt}

\autoref{fig:prompt} reproduces the prompt template used to elicit intent labels for each adjacent pair (\autoref{sec:intent_analysis}). The placeholders \keyword{\_\_COMMIT\_A\_\_} and \keyword{\_\_COMMIT\_B\_\_} are replaced with the rule text, and the model is queried in JSON-only structured-output mode with the schema shown. The body text uses shortened names \op{added}, \op{removed}, \op{modified} for the JSON fields \keyword{predicate\_added}, \keyword{predicate\_removed}, \keyword{predicate\_modified\_present}, respectively.

The schema additionally collects \keyword{summary}, \keyword{rationale\_confidence}, and \keyword{rationale\_support} fields; these are retained for manual spot-checking and error analysis but are not used in any aggregate analysis reported in this paper.

\begin{figure*}[p]
\begin{lstlisting}[
basicstyle=\sffamily\small,
breaklines=true,
  breakatwhitespace=false,
  columns=flexible,
breakindent=0em,
breaklines=true]

You are analyzing predicate-level logic changes between two detection rule versions. Predicate logic is known to have changed. Characterize it using the schema below.

## Output (strict JSON)
{
  "from_commit": "<hash>",
  "to_commit": "<hash>",
  "match_set_direction": "broader | narrower | mixed | unclear",
  "predicate_modified_present": bool,
  "predicate_added": bool,
  "predicate_removed": bool,
  "summary": "<one sentence: observable logic change>",
  "rationale_label": "coverage_expansion | false_positive_reduction | mixed_tradeoff | insufficient_evidence",
  "rationale_confidence": "high | medium | low",
  "rationale_support": "<brief explanation grounded only in the observed pair>"
}
## Definitions
- predicate_modified_present: An existing predicate is rewritten in-place (field, operator, or value changed)
- predicate_added: Any new predicate introduced into the logic (including AND constraints or OR branches)
- predicate_removed: Any predicate removed from the logic (including AND constraints or OR branches)
## Rules
- Analyze predicate logic only (conditions, values, Boolean structure)
- Ignore formatting, macros, output fields, and pipeline mechanics
- Do not infer intent beyond the pair
- match_set_direction: broader = matches more; narrower = matches fewer; mixed = both; unclear = cannot determine
- rationale_label: be conservative; use insufficient_evidence if intent is ambiguous
## Example
Commit A:
(CommandLine contains whoami/systeminfo/&cd&echo) OR (CommandLine contains net AND user) OR (CommandLine contains cd AND /d) OR (CommandLine contains ping AND -n)
Commit B:
(Image endswith whoami.exe/systeminfo.exe) OR (Image endswith net.exe/net1.exe AND CommandLine contains user) OR (CommandLine contains cd AND /d) OR (Image endswith ping.exe AND CommandLine contains -n) OR (CommandLine contains &cd&echo)
Output:
{
  "from_commit": "A",
  "to_commit": "B",
  "match_set_direction": "mixed",
  "predicate_modified_present": true,
  "predicate_added": true,
  "predicate_removed": true,
  "summary": "Rewrites command-line predicates into executable-based conditions, adds new constraints, and introduces additional alternatives.",
  "rationale_label": "mixed_tradeoff",
  "rationale_confidence": "medium",
  "rationale_support": "The rule replaces broad command-line checks with more specific executable-based conditions while adding new alternatives."
}
## Now analyze:
### Commit A
__COMMIT_A__
### Commit B
__COMMIT_B__
\end{lstlisting}
\caption{LLM Prompt for obtaining intent labels}\label{fig:prompt}
\end{figure*}

\ifCAMERAREADY
\else
\section{Ethical Considerations}  
This work analyzes the commit histories of two public, openly licensed detection rule repositories: the community-driven Sigma project and Splunk Security Content. No human subjects, telemetry, enterprise deployment data, or non-public artifacts were collected or analyzed.
All authorship and revision metadata used in our analysis is part of the public git history of these repositories. We did not contact rule authors, attempt to deanonymize contributors, or aggregate contributor-level information; our unit of analysis throughout is the rule lineage, not the developer. For these reasons, IRB review was not required, and no contributor or end-user data is at risk of re-identification through our pipeline or released artifacts.

\shortsection{Dual-use considerations}
Detection rules in public repositories are, by design, already visible to both defenders and adversaries. Our analysis does not expose previously private rules, propose new evasion techniques, or identify specific detection gaps that are not already apparent from reading the rules themselves. The contributions of this paper operate at a higher level of abstraction than any individual rule: we characterize \emph{how rules evolve} (e.g., the prevalence of paired expansion and exclusion, or of structural reversion) rather than \emph{which rules are weak}. 
We considered whether aggregate evolution patterns could meaningfully aid evasion and concluded that they do not, because the structural and intent patterns we surface (coverage expansion, false-positive reduction, mixed tradeoffs) describe the maintenance dynamics of detection engineering as a discipline rather than localized bypass surfaces. We believe the benefit to the defender community---empirical grounding for evaluation frameworks, AI-assisted rule authoring, and tooling that operates over canonicalized rule semantics---substantially outweighs this residual risk.
\fi

%% file: meta_review.tex
\section{Ethical Considerations}  
\shortsection{
Ethical Considerations for Human Subjects Research}
This work analyzes the commit histories of two public, openly licensed detection rule repositories: the community-driven Sigma project and Splunk Security Content. No human subjects, telemetry, enterprise deployment data, or non-public artifacts were collected or analyzed.

All authorship and revision metadata used in our analysis is part of the public git history of these repositories. We did not contact rule authors, attempt to deanonymize contributors, or aggregate contributor-level information; our unit of analysis throughout is the rule lineage, not the developer. For these reasons, IRB review was not required, and no contributor or end-user data is at risk of re-identification through our pipeline or released artifacts.

\shortsection{Ethics Considerations for Developing New Tools or Technologies}
Detection rules in public repositories are, by design, already visible to both defenders and adversaries. Our analysis does not expose previously private rules, propose new evasion techniques, or identify specific detection gaps that are not already apparent from reading the rules themselves. The contributions of this paper operate at a higher level of abstraction than any individual rule: we characterize how rules evolve (e.g., the prevalence of paired expansion and exclusion, or of structural reversion) rather than which rules are weak.

We considered whether aggregate evolution patterns could meaningfully aid evasion and concluded that they do not, because the structural and intent patterns we surface (coverage expansion, false-positive reduction, mixed tradeoffs) describe the maintenance dynamics of detection engineering as a discipline rather than localized bypass surfaces. We believe the benefit to the defender community---empirical grounding for evaluation frameworks, AI-assisted rule authoring, and tooling that operates over canonicalized rule semantics---substantially outweighs this residual risk.

\section{Meta-Review}

The following meta-review was prepared by the program committee for the 2027
IEEE Symposium on Security and Privacy (S\&P) as part of the review process as
detailed in the call for papers.

\subsection{Summary}
The paper proposes a predicate-graph-based approach for a longitudinal study of structural changes in host-based intrusion detection rules. Graph-based analysis identifies these changes, while LLM inference establishes their underlying purpose. The evaluation is performed on the Sigma project and Splunk's Security Content rules over their entire lifespan, highlighting behavioral patterns in the maintainer communities.

\subsection{Scientific Contributions}
\begin{enumerate}
    \item Addresses a Long-Known Issue
    \item Independent Confirmation of Important Results with Limited Prior Research
    \item Provides a Valuable Step Forward in an Established Field
\end{enumerate}

\subsection{Reasons for Acceptance}
\begin{enumerate}
    \item Addresses a Long-Known Issue. Inconsistencies in HIDS rules have been widely explored structurally, this study extends that exploration to rule semantics.
    \item Independent Confirmation of Important Results with Limited Prior Research. The cross-checking structural and natural language analysis provides a useful blueprint for future studies, and demonstrates that LLMs are largely accurate for this analysis task.
    \item Provides a Valuable Step Forward in an Established Field. Analysis of structural rule evolution reveals not only how rules evolve but, more importantly, why. These observations provide insight into operational trade-offs and, consequently, how to build security rules better.
\end{enumerate}

\subsection{Noteworthy Concerns} 
\begin{enumerate} 
    \item The framing of the implications of the results for security could be stronger. It remains unclear whether certain constructs for rule writing are more prone to errors than others and what guidelines might improve rule writing in terms of speed, consistency, and fewer false positives.
\end{enumerate}

\section{Response to the Meta-Review} 

The meta-review notes that we do not provide specific guidance in the paper as to how to improve rule writing. We agree with this point, but would like to provide context on why we deliberately stop short of such guidance.

Our study measures detection-logic evolution from public rule histories. These histories record what changed, when, and in what structural form, but they do not record how a rule behaved in deployment. Establishing that a construct is error-prone requires linking it to outcomes such as alert volume, false-positive rates, analyst dispositions, and missed detections, none of which are directly observable in the Sigma or SSC repositories. False positives are a property of a rule in an environment, not of a rule alone. A predicate that is precise in one enterprise may be unusably noisy in another with different log sources, baseline administrative behavior, endpoint software, or alert-triage capacity.

Guidelines derived only from public repository activity would therefore risk being presented as general when they are in fact environment-specific. Hence, we limit our conclusions to reporting what our measurements establish and are reluctant to offer guidance whose validity we cannot assess. Our work provides a method that makes predicate- and scope-level change measurable at corpus scale. That instrument is directly applicable to private rule histories inside vendors and enterprises, where revisions can be joined to deployment feedback. Applied there, the questions the meta-review raises become answerable empirically, and the answers can be calibrated to the environment in which they will be used.